\documentclass{article}%
\usepackage{amsmath}
\usepackage{amsfonts}
\usepackage{amssymb}
\usepackage{graphicx}%
\providecommand{\U}[1]{\protect\rule{.1in}{.1in}}
\newtheorem{theorem}{Theorem}

\newtheorem{definition}[theorem]{Definition}

\newtheorem{exercise}[theorem]{Exercise}

\newtheorem{remark}[theorem]{Remark}
\newtheorem{solution}[theorem]{Solution}

\begin{document}

\DeclareGraphicsExtensions{.eps}

\title{Extracting Energy from an External Magnetic Field}
\author{Waldyr Alves Rodrigues Jr. and Edmundo Capelas de Oliveira\thanks{email:
capelas@ime.unicamp.br}\\Institute of Mathematics, Statistics and Scientific Computation\\IMECC-UNICAMP\\email: walrod@ime.unicamp.br \ or walrod@mpc.com.br}
\maketitle

\begin{abstract}
In this paper we describe the theory of a device that is able to extract
energy from an external magnetic field. The device is a cylindrical magnetic
insulator that once put in rotation makes electromagnetic angular momentum to
be stored in the electromagnetic field in contrary direction to the mechanical
angular momentum of the device. As a consequence due to total angular momentum
conservation the device increases its angular velocity (when
$\mathbf{\varepsilon\mu}>1$)\footnote{Natural units are used in the paper.}
and all conservation laws are rigorously satisfied. \ The voltage generated by
the device is found solving\ explicitly Maxwell equations for rotating
magnetic insulators in external fields a subject that have provoked lots of
polemics in the literature and which we hope to be here clarified due to our
pedagogical presentation.

\end{abstract}

\section{Introduction}

In this paper we introduce the theory of a device that can in principle
extract energy from an external magnetic field. The idea for the device came
to us from the theoretical analysis of the 1913 Wilson \& Wilson \cite{ww}
experiment (WWE)\footnote{The experiment has been repeated by Hertzberg and
collaborators \cite{her} in 2001.} described in Section 5. Since the correct
theoretical explanation of the experiment gave rise to
controversies\footnote{A sample may be found in
\cite{pelsw,rid,webster,webster1,shio} and the inumerous references to related
issues in those papers.} in the literature due to difficulties in properly
applying Maxwell theory in rotating frames\footnote{See, e.g.,\ the most
quoted Schiff's paper \cite{schiff} where it is (wrongly) claimed that one
need use General Relativity to explain electrodynamics phenomena in rotating
frames. Besides that let us recall the almost continuous difficulty that some
have in each generation to understand simple electromagnetic phenomena when
moving boundaries are present, as e.g., in the application of Faraday's law of
induction in the homopolar generator. A thoughtful discussion of the
application of Faraday's law in its many equivalent formulations may be found
in \cite{rodfb}.} we decided to give a very pedagogical presentation for the
problem of expressing Maxwell equations in a coordinate invariant way using
the theory of differential forms. Then, using that formalism and the correct
jumping conditions for the electromagnetic field variables between moving
boundaries separating two different media we get in a very clear way the
electromagnetic field as measured in the laboratory in the
WWE.\footnote{Application of differential forms formalism to the WWE already
appeared in \cite{catu,ho}. Our approach details each step in the derivation
and hopefuly is a little bit more pedagogical.} We think that our approach
clears up immediately what is wrong with some other attempts to find a correct
description of the WWE in the sample of papers quoted. The description of the
machine to extract energy from an external magnetic field is given in section
6 where it is observed that the machine once puts in motion makes
electromagnetic angular to be stocked in the electric plus magnetic fields in
the opposite direction of the mechanical angular mechanical of the device
(when $\mathbf{\mu\varepsilon}>1$) and thus due to angular momentum
conservation the machine has its mechanical angular momentum increased. Of
course, all the conservation laws are in operation in our device and the
energy being generated comes from the external electromagnetic field.

\section{Maxwell Equations}

The formulation of Maxwell equations in intrinsic form requires as preliminary
mathematical structures:

(i) a $4$-dimensional orientable\footnote{By orientable we mean that there
exists in $M$ a global $4$-form $\boldsymbol{\tau}_{\boldsymbol{g}}\in\sec%
{\textstyle\bigwedge^{4}}
T^{\ast}M$. If the manifold is not orientable then it is necessary to use
besides the concepts of \emph{pair }form fields also the concepts of
\emph{impair} form fields. Details may be found in \cite{ho}. See also a
thoughtful discussion in \cite{roro2010}.} manifold $M$, the bundle of non
homogeneous differential forms $%
{\textstyle\bigwedge}
T^{\ast}M=%
{\textstyle\bigoplus\nolimits_{r=0}^{r=4}}
{\textstyle\bigwedge^{r}}
T^{\ast}M$, where $%
{\textstyle\bigwedge^{r}}
T^{\ast}M$ is the bundle of $r$-forms and

(ii) the differential operator $d:%
{\textstyle\bigwedge^{r}}
T^{\ast}M\rightarrow%
{\textstyle\bigwedge^{r+1}}
T^{\ast}M$, $d^{2}=0$.

Indeed, Maxwell equations deals with a field $F\in\sec%
{\textstyle\bigwedge^{2}}
T^{\ast}M$ which is exact, i.e., $F=dA$, where $A\in\sec%
{\textstyle\bigwedge^{1}}
T^{\ast}M$ and a current $\mathbf{J}\in\sec%
{\textstyle\bigwedge^{3}}
T^{\ast}M$ that is also exact, i.e., $\mathbf{J}=dH$, where $H\in\sec%
{\textstyle\bigwedge^{2}}
T^{\ast}M$. The set of equations
\begin{equation}
dF=0,\text{ \ \ }dH=-\mathbf{J}, \label{m1}%
\end{equation}
is known as Maxwell equations and of course, we have
\begin{equation}
d\mathbf{J}=0, \label{m2}%
\end{equation}
i.e., the current is conserved.

Maxwell equations are \emph{invariant under diffeomorphisms} as it happens
with all theories formulated with differential forms. This means that if
$h:M\rightarrow M$ is a diffeomorphism, then denoting as usual the pullback
mapping by $h^{\ast}$, since the differential $d$ commutes with the pullback,
i.e., $dh^{\ast}=h^{\ast}d$ we have that the fields $F^{\prime}=h^{\ast
}F,H^{\prime}=h^{\ast}H,\mathbf{J}^{\prime}=h^{\ast}\mathbf{J}$ satisfy%

\begin{equation}
dF^{\prime}=0,\text{ \ \ }dH^{\prime}=-\mathbf{J}^{\prime}. \label{m3}%
\end{equation}

Note that we did not make until now any requirements concerning the topology
of manifold $M$, so to proceed we take \emph{arbitrary coordinates} $\langle
x^{\mu}\rangle$ covering $U\subset M$ and denoting\footnote{Take notice that
coordinate vector fields are denoted using a bold symbol $\boldsymbol{\partial
}$, e.g., $\boldsymbol{\partial/\partial}x^{\mu}=\boldsymbol{\partial}_{\mu}$,
whereas we use the symbol $\partial/\partial x^{\mu}\boldsymbol{:}%
=\partial_{\mu}$ to denote the usual partial derivatives. This means the
following: Let $f:M\rightarrow\mathbb{R}$ a differentiable function and let
$(U,\psi$) be a chart of the atlas of $M$ such that for $e\in U$,
$\psi(e)=(x^{0},x^{1},x^{2},x^{3})$. Then, $\frac{\boldsymbol{\partial}%
}{\boldsymbol{\partial}x^{\mu}}f=\frac{\partial}{\partial x^{\mu}}f\circ\psi
$.} $\langle e_{\mu}=\boldsymbol{\partial}/\boldsymbol{\partial}x^{\mu
}=\boldsymbol{\partial}_{\mu}\rangle$ the corresponding coordinate basis for
$TU$ and by $\langle\vartheta^{\mu}=dx^{\mu}\rangle$ the basis for $T^{\ast}U$
dual to the basis $\langle e_{\mu}\rangle$ we write
\begin{align}
F  &  =\frac{1}{2}F_{\mu\nu}\vartheta^{\mu}\wedge\vartheta^{\nu},\nonumber\\
H  &  =\frac{1}{2}H_{\mu\nu}\vartheta^{\mu}\wedge\vartheta^{\nu},\label{m4}\\
\mathbf{J}  &  =\frac{1}{3!}J_{\mu\nu\rho}\vartheta^{\mu}\wedge\vartheta^{\nu
}\boldsymbol{\wedge}\vartheta^{\nu}.\nonumber
\end{align}

Maxwell equations are supposed to describe the behavior of electromagnetic
fields in vacuum or in material media. But whereas, according to Feynman, the
field $F$ is fundamental (it define the Lorentz force acting on probe charged
particles moving in the field) the field $H$ is phenomenological\footnote{At
least at the classical level. This is so because the calculation of $H$ needs
in the general case sophisticated use of quantum theory.}. In fact, if we
write Fourier representation for $F$ and $H,$%
\begin{align}
F_{\mu\nu}(x)  &  =%
{\textstyle\int}
d^{4}k\breve{F}_{\mu\nu}(k)e^{-i(k_{0}x^{0}+k_{1}x^{1}+k_{2}x^{2}+k_{3}x^{3}%
)}:=%
{\textstyle\int}
d^{4}k\mathfrak{F}_{\mu\nu}(k,x),\nonumber\\
H_{\mu\nu}(x)  &  =%
{\textstyle\int}
d^{4}k\text{ }\breve{H}_{\mu\nu}(k)e^{-i(k_{0}x^{0}+k_{1}x^{1}+k_{2}%
x^{2}+k_{3}x^{3})}:=%
{\textstyle\int}
d^{4}k\mathfrak{H}_{\mu\nu}(k,x), \label{m5}%
\end{align}
in general we have, defining the 2-form valued function $\boldsymbol{\breve
{\varkappa}}:\mathbb{R\times}\sec%
{\textstyle\bigwedge^{2}}
T^{\ast}M\rightarrow\sec%
{\textstyle\bigwedge^{2}}
T^{\ast}M$ (the frequency constituent equations of the medium) that%
\begin{equation}
\mathfrak{H}(k_{0},x)=\boldsymbol{\breve{\varkappa}}(\mathfrak{F}(k_{0},x)).
\label{M5}%
\end{equation}

The function $\boldsymbol{\breve{\varkappa}}$ may even be in some media a
nonlinear function of $\mathfrak{F}$ (see, e.g., \cite{jackson}) but in what
follows we will consider only non dispersive media ($\boldsymbol{\breve
{\varkappa}}$ is independent of $k_{0}$) in which case we can define a linear
constituent function (an extensor field)
\begin{gather}
\boldsymbol{\varkappa}:\sec%
{\textstyle\bigwedge\nolimits^{2}}
T^{\ast}M\rightarrow\sec%
{\textstyle\bigwedge\nolimits^{2}}
T^{\ast}M,\nonumber\\
H=\boldsymbol{\varkappa}(F). \label{m6}%
\end{gather}
In this case we have for the components $H_{\mu\nu}$ of $H$,%
\begin{equation}
H_{\mu\nu}=\frac{1}{2}\boldsymbol{\varkappa}_{\mu\nu\cdot\cdot}^{\cdot
\cdot\alpha\beta}F_{\alpha\beta}. \label{m7}%
\end{equation}

Of course, we have the obvious symmetries for the components of the
constituent extensor\footnote{We recall that these are the same symmetries of
the components of the Riemann tensor of a metric compatible connection.}
$\boldsymbol{\varkappa}$,
\begin{equation}
\boldsymbol{\varkappa}_{\mu\nu\cdot\cdot}^{\cdot\cdot\alpha\beta
}=-\boldsymbol{\varkappa}_{\nu\mu\cdot\cdot}^{\cdot\cdot\alpha\beta},\text{
\ \ }\boldsymbol{\varkappa}_{\mu\nu\cdot\cdot}^{\cdot\cdot\alpha\beta
}=-\boldsymbol{\varkappa}_{\mu\nu\cdot\cdot}^{\cdot\cdot\beta\alpha},\text{
\ \ }\boldsymbol{\varkappa}_{\mu\nu\cdot\cdot}^{\cdot\cdot\alpha\beta
}=\boldsymbol{\varkappa}_{\nu\mu\cdot\cdot}^{\cdot\cdot\beta\alpha}\text{ }.
\label{m8}%
\end{equation}

\subsection{Enter $\langle M,\boldsymbol{g},\mathbf{\nabla},\boldsymbol{\tau
}_{\boldsymbol{g}},\uparrow\rangle$}

We can proceed with the formulation of Maxwell theory without the introduction
of additional mathematical objects in the manifold $M$. If you are interested
in knowing the details, please consult \cite{ho}. From now on we will suppose
that electromagnetic fields are to be described in a Lorentzian spacetime
structure $\langle M,\boldsymbol{g},\mathbf{\nabla},\boldsymbol{\tau
}_{\boldsymbol{g}},\uparrow\rangle$, where $\langle M,\boldsymbol{g}%
\boldsymbol{\rangle}$ is a Lorentzian manifold, $\mathbf{\nabla}$ is the
Levi-Civita connection of $\boldsymbol{g}$, $\boldsymbol{\tau}_{\boldsymbol{g}%
}=\sqrt{\left\vert \det\boldsymbol{g}\right\vert }\boldsymbol{\vartheta}%
^{0}\wedge\boldsymbol{\vartheta}^{1}\wedge\boldsymbol{\vartheta}^{2}%
\wedge\boldsymbol{\vartheta}^{3}$ is the metrical volume element and
$\uparrow$ denotes that the structure $\langle M,\boldsymbol{g)}$ is time orientable.

\begin{remark}
As it is well known a spacetime structure $(M,\boldsymbol{g},\mathbf{\nabla
},\boldsymbol{\tau}_{\boldsymbol{g}},\uparrow\rangle$ when the Riemann
curvature tensor of $\mathbf{\nabla}$ is non null represents a gravitational
field generated by an energy-momentum tensor $T$ in Einstein's General
Relativity \emph{(}GR\emph{)} where Einstein equation is satisfied. It is a
simple exercise to find an effective spacetime structure, say $\langle
M,\mathbf{g},\mathbf{\nabla},\boldsymbol{\tau}_{\mathbf{g}},\uparrow\rangle$,
to describe some material media. However, such structure has nothing to do
with GR.
\end{remark}

\subsubsection{Reference Frames}

We shall need in what follows the concepts of reference frames, observers and
naturally adapted coordinate systems to a reference frame in a general
Lorentzian spacetime structure $\langle M,\boldsymbol{g},\mathbf{\nabla
},\boldsymbol{\tau}_{\boldsymbol{g}},\uparrow\rangle$.

We define a \emph{reference frame} in $U\subset M$ as a\ unit timelike vector
field $\boldsymbol{Z}$ pointing to the future. We have
\begin{equation}
\boldsymbol{g}(\boldsymbol{Z},\boldsymbol{Z})=1. \label{rf1}%
\end{equation}

An \emph{observer} is defined as a timelike curve $\sigma$ in $M$
(parametrized with proper time) and pointing to the future. We denote by
$\sigma_{\ast s}$ the tangent vector at $\sigma(s)$ and%
\begin{equation}
\boldsymbol{g}(\sigma_{\ast s},\sigma_{\ast s})=1. \label{rf2}%
\end{equation}
We immediately realize that each one of the integral lines of $\boldsymbol{Z}%
$, is an observer.

Finally we say that coordinates $\langle x^{\mu}\rangle$ covering $U\subset M$
is a \emph{naturally adapted coordinate system to a reference frame
}$\boldsymbol{Z}$ (nacs%
$\vert$%
$\boldsymbol{Z}$) if
\begin{equation}
\boldsymbol{Z}=\frac{1}{\sqrt{g_{00}}}\frac{\boldsymbol{\partial}%
}{\boldsymbol{\partial}x^{0}}\text{ }. \label{rf3}%
\end{equation}

A detailed classification of reference frames in a Lorentzian spacetime may be
found, e.g., in \cite{rodcap2007}. For a classification of reference frames in
a Riemann-Cartan spacetime, see, e.g., \cite{gr2012}.

\subsubsection{Formulation of Vacuum Maxwell Equations in $\langle
M,\boldsymbol{g},\mathbf{\nabla},\boldsymbol{\tau}_{\boldsymbol{g}}%
,\uparrow\rangle$}

Given the structure $\langle M,\boldsymbol{g},\mathbf{\nabla},\boldsymbol{\tau
}_{\boldsymbol{g}},\uparrow\rangle$\ it is convenient to define
\begin{equation}
G=\underset{\boldsymbol{g}}{\star}^{-1}H
\end{equation}
where $\underset{\boldsymbol{g}}{\star}:\sec%
{\textstyle\bigwedge\nolimits^{r}}
T^{\ast}M\rightarrow\sec%
{\textstyle\bigwedge\nolimits^{4-r}}
T^{\ast}M$ is the Hodge star operator and $\underset{\boldsymbol{g}}{\star
}^{-1}\underset{\boldsymbol{g}}{\star}=\underset{\boldsymbol{g}}{\star
}\underset{\boldsymbol{g}}{\star}^{-1}=\mathrm{Id}$. It is also a standard
practice to introduce $J\in\sec%
{\textstyle\bigwedge\nolimits^{1}}
T^{\ast}M$ such that \
\begin{equation}
\mathbf{J=}\underset{\boldsymbol{g}}{\star}J. \label{m9a}%
\end{equation}
We define also the (ex)tensor field
\begin{gather}
\boldsymbol{\chi}:\sec%
{\textstyle\bigwedge\nolimits^{2}}
T^{\ast}M\rightarrow\sec%
{\textstyle\bigwedge\nolimits^{2}}
T^{\ast}M,\label{m9aa}\\
G=\boldsymbol{\chi}(F).\nonumber
\end{gather}
Writing\footnote{$\langle\vartheta_{\mu}\rangle$ is the reciprocal basis of
$\langle\vartheta^{\mu}\rangle$, i.e., \ $\mathtt{g}(\vartheta^{\mu}%
,\vartheta_{\nu})=\delta_{\nu}^{\mu}$.}
\begin{equation}
G=\frac{1}{2}G_{\mu\nu}\vartheta^{\mu}\wedge\vartheta^{\nu}=\frac{1}{2}%
G^{\mu\nu}\vartheta_{\mu}\wedge\vartheta_{\nu} \label{m9ab}%
\end{equation}
we have%
\begin{equation}
G_{\mu\nu}=\frac{1}{2}\chi_{\mu\nu\cdot\cdot}^{\cdot\cdot\alpha\beta}%
F_{\alpha\beta},\text{ \ \ }G^{\mu\nu}=\frac{1}{2}\chi^{\mu\nu\alpha\beta
}F_{\alpha\beta} \label{m9ac}%
\end{equation}
and of course, the components of the constituent extensor $\boldsymbol{\chi}$
satisfy
\begin{equation}
\chi^{\mu\nu\alpha\beta}=-\chi^{\nu\mu\alpha\beta},\text{ \ \ }\chi^{\mu
\nu\alpha\beta}=-\chi^{\mu\nu\beta\alpha},\text{ \ \ }\chi^{\mu\nu\alpha\beta
}=\chi^{\nu\mu\beta\alpha}. \label{m9ad}%
\end{equation}
\ 

In vacuum the relation between the fields $G$ and $F$ is%
\begin{equation}
G=F \label{m9}%
\end{equation}
and we write Maxwell equations as
\begin{equation}
dF=0,\text{ \ \ \ \ }d\underset{\boldsymbol{g}}{\star}%
F=-\underset{\boldsymbol{g}}{\star}J. \label{m9b}%
\end{equation}
or better yet, applying the inverse of the Hodge star operator to both members
of the non homogenous equation we can write%
\[
dF=0,\text{ \ \ \ \ }\underset{\boldsymbol{g}}{\delta}F=-J
\]
We recall that\footnote{$\boldsymbol{\vartheta}^{\mu_{1}...\mu_{p}%
}:=\boldsymbol{\vartheta}^{\mu_{1}...\mu_{p}}\wedge\boldsymbol{\vartheta}%
^{\mu_{2}}\wedge\cdots\wedge\boldsymbol{\vartheta}^{\mu_{p}}.$}%

\begin{equation}
\underset{\boldsymbol{g}}{\star}\vartheta^{\mu_{1}...\mu_{p}}=\frac{1}%
{(n-p)!}\sqrt{\left\vert \det\boldsymbol{g}\right\vert }g^{\mu_{1}\nu_{1}%
}...g^{\mu_{p}\nu_{p}}\epsilon_{\nu_{1}...\nu_{n}}\vartheta^{\nu_{p+1}%
...\nu_{n}}. \label{m10}%
\end{equation}

Then, write
\begin{equation}
H=\underset{\boldsymbol{g}}{\star}F=\frac{1}{2}H_{\rho\sigma}\vartheta^{\rho
}\wedge\vartheta^{\sigma}=\frac{1}{2}H^{\rho\sigma}\vartheta_{\rho}%
\wedge\vartheta_{\sigma} \label{m11}%
\end{equation}
and let us calculate $\underset{\boldsymbol{g}}{\star}F$. We have%

\begin{align}
\underset{\boldsymbol{g}}{\star}F  &  =\frac{1}{2}F_{\mu\nu}%
\underset{\boldsymbol{g}}{\star}(\vartheta^{\mu}\wedge\vartheta^{\nu
})\nonumber\\
&  =\frac{1}{2}\frac{1}{2}F_{\mu\nu}\sqrt{\left\vert \det\boldsymbol{g}%
\right\vert }g^{\mu\alpha}g^{\nu\beta}\epsilon_{\alpha\beta\rho\sigma
}\vartheta^{\rho}\wedge\vartheta^{\sigma}. \label{m12}%
\end{align}

From Eq.(\ref{m11}) and Eq.(\ref{m12}) we have%
\begin{align}
H_{\rho\sigma}  &  =\frac{1}{2}\sqrt{\left\vert \det\boldsymbol{g}\right\vert
}g^{\mu\alpha}g^{\nu\beta}\epsilon_{\alpha\beta\rho\sigma}F_{\mu\nu
}\label{m13}\\
&  =\frac{1}{2}\sqrt{\left\vert \det\boldsymbol{g}\right\vert }g^{\mu\alpha
}g^{\nu\beta}\epsilon_{\rho\sigma\alpha\beta}F_{\mu\nu}\nonumber\\
&  =\frac{1}{2}\sqrt{\left\vert \det\boldsymbol{g}\right\vert }\epsilon
_{\rho\sigma\cdot\cdot}^{\cdot\cdot\mu\nu}F_{\mu\nu},\nonumber
\end{align}
or taking into account that
\begin{equation}
H_{\rho\sigma}=\text{ }^{\ast}G_{\rho\sigma}=\frac{1}{2}\sqrt{\left\vert
\det\boldsymbol{g}\right\vert }\epsilon_{\rho\sigma\cdot\cdot}^{\cdot\cdot
\mu\nu}G_{\mu\nu}%
\end{equation}
we get%
\begin{equation}
G^{\rho\sigma}=g^{\rho\mu}g^{\sigma\nu}F_{\mu\nu}=\frac{1}{2}(g^{\rho\mu
}g^{\sigma\nu}-g^{\rho\nu}g^{\sigma\mu})F_{\mu\nu}.
\end{equation}

Comparing this equation with Eq.(\ref{m9ac}) gives for the\ components
$\varkappa^{\rho\sigma\mu\nu}$ of the constitutive (ex)tensor of the vacuum%
\begin{equation}
\chi^{\rho\sigma\mu\nu}=(g^{\rho\mu}g^{\sigma\nu}-g^{\rho\nu}g^{\sigma\mu}).
\label{m17}%
\end{equation}

\begin{remark}
Before proceeding let us recall that the similarity between the components of
$\chi^{\rho\sigma\mu\nu}$ and the components $\boldsymbol{R}^{\rho\sigma\mu
\nu}$ of a Riemann curvature tensor $\boldsymbol{R}$ with a
$\boldsymbol{constant}$ Riemann curvature scalar $K$. Indeed, we have
\begin{equation}
\boldsymbol{R}^{\rho\sigma\mu\nu}=\frac{K}{12}(g^{\rho\mu}g^{\sigma\nu
}-g^{\rho\nu}g^{\sigma\mu}) \label{m18}%
\end{equation}
and thus $K=R=\delta_{\beta}^{\nu}R_{\cdot\cdot\alpha\nu}^{\alpha\beta
\cdot\cdot}$. Take notice that $\boldsymbol{R}$ is not the Riemann curvature
tensor of the Levi-Civita connection $D$ of $\boldsymbol{g}$.
\end{remark}

\begin{exercise}
Show that in components Maxwell\ equations $dF=0$,
$dH=-\underset{\boldsymbol{g}}{\star}J$ \ when $\boldsymbol{\varkappa}$ has
its vacuum value read%
\begin{align}
\partial_{\mu}F_{\alpha\beta}+\partial_{\beta}F_{\mu\alpha}+\partial_{\alpha
}F_{\beta\mu}  &  =0,\label{exercise0}\\
\frac{1}{\sqrt{\left\vert \det\boldsymbol{g}\right\vert }}\partial_{\mu}%
(\sqrt{\left\vert \det\boldsymbol{g}\right\vert }F^{\mu\nu})  &  =J^{\nu}.
\label{exercise}%
\end{align}

\end{exercise}

\begin{solution}
We have%
\begin{align*}
dF  &  =\frac{1}{2}d(F_{\alpha\beta}\wedge\vartheta^{\alpha}\wedge
\vartheta^{\beta})\\
&  =\frac{1}{2}\partial_{\mu}F_{\alpha\beta}\vartheta^{\mu}\wedge
\vartheta^{\alpha}\wedge\vartheta^{\beta}\\
&  =\frac{1}{3!}(\partial_{\mu}F_{\alpha\beta}+\partial_{\beta}F_{\mu\alpha
}+\partial_{\alpha}F_{\beta\mu})\vartheta^{\mu}\wedge\vartheta^{\alpha}%
\wedge\vartheta^{\beta}%
\end{align*}
and \emph{Eq.(\ref{exercise0}) }follows\emph{.}

Next we define $J=J_{\mu}\vartheta^{\mu}:=\rho\vartheta^{0}-j_{i}\vartheta
^{i}$ and get%
\begin{equation}
\underset{\boldsymbol{g}}{\star}J=\frac{1}{3!}\sqrt{\left\vert \det
\boldsymbol{g}\right\vert }J_{\lambda}g^{\lambda\kappa}\epsilon_{\kappa
\mu\alpha\beta}\vartheta^{\mu}\wedge\vartheta^{\alpha}\wedge\vartheta^{\beta}.
\label{ex1}%
\end{equation}
Then the equation $dH=-\underset{\boldsymbol{g}}{\star}J$ reads%
\begin{equation}
\frac{1}{3!}(\partial_{\mu}H_{\alpha\beta}+\partial_{\beta}H_{\mu\alpha
}+\partial_{\alpha}H_{\beta\mu})\vartheta^{\mu}\wedge\vartheta^{\alpha}%
\wedge\vartheta^{\beta}=-\frac{1}{3!}\sqrt{\left\vert \det\boldsymbol{g}%
\right\vert }J^{\kappa}\epsilon_{\kappa\mu\alpha\beta}\vartheta^{\mu}%
\wedge\vartheta^{\alpha}\wedge\vartheta^{\beta} \label{ex2}%
\end{equation}
and
\begin{equation}
(\partial_{\mu}H_{\alpha\beta}+\partial_{\beta}H_{\mu\alpha}+\partial_{\alpha
}H_{\beta\mu})=-\sqrt{\left\vert \det\boldsymbol{g}\right\vert }J^{\kappa
}\epsilon_{\kappa\mu\alpha\beta}. \label{ex3}%
\end{equation}
Multiplying both members of the last equation by $\epsilon^{_{\lambda\mu
\alpha\beta}}$ and recalling\footnote{See, e.g. page 111 of \cite{loverund}.}
that
\begin{align}
\epsilon^{_{\lambda\mu\alpha\beta}}\epsilon_{\kappa\mu\alpha\beta}  &
=3!\delta_{\kappa}^{\lambda}\label{ex4}\\
\epsilon^{_{\lambda\mu\alpha\beta}}\epsilon_{\alpha\beta\kappa\iota}  &
=2!\delta_{\kappa\iota}^{\lambda\mu}\nonumber
\end{align}
we have%
\begin{equation}
\epsilon^{_{\lambda\mu\alpha\beta}}(\partial_{\mu}H_{\alpha\beta}%
+\partial_{\beta}H_{\mu\alpha}+\partial_{\alpha}H_{\beta\mu})=-3!\sqrt
{\left\vert \det\boldsymbol{g}\right\vert }J^{\lambda}. \label{EX5}%
\end{equation}
Now,
\begin{align}
\epsilon^{_{\lambda\mu\alpha\beta}}\partial_{\mu}H_{\alpha\beta}  &
=\partial_{\mu}(\frac{1}{2}\sqrt{\left\vert \det\boldsymbol{g}\right\vert
}\epsilon^{_{\lambda\mu\alpha\beta}}\epsilon_{\alpha\beta\kappa\iota}%
F^{\kappa\iota})\nonumber\\
&  =\partial_{\mu}(\sqrt{\left\vert \det\boldsymbol{g}\right\vert }%
\delta_{\kappa\iota}^{\lambda\mu}F^{\kappa\iota})\nonumber\\
&  =2\partial_{\mu}(\sqrt{\left\vert \det\boldsymbol{g}\right\vert }%
F^{\lambda\mu})\nonumber\\
&  =-2\partial_{\mu}(\sqrt{\left\vert \det\boldsymbol{g}\right\vert }%
F^{\mu\lambda}). \label{ex6}%
\end{align}
Analogous calculations give that the other two terms in the first member of
\emph{Eq.(\ref{EX5})} are also equal to $-2\partial_{\mu}(\sqrt{\left\vert
\det\boldsymbol{g}\right\vert }F^{\mu\lambda})$. Then \emph{Eq.(\ref{EX5})}
becomes%
\[
\frac{1}{\sqrt{\left\vert \det\boldsymbol{g}\right\vert }}\partial_{\mu}%
(\sqrt{\left\vert \det\boldsymbol{g}\right\vert }F^{\mu\lambda})=J^{\lambda}%
\]
and \emph{Eq.(\ref{exercise})} is proved.
\end{solution}

\begin{remark}
\label{clifford}We present yet another simple proof of
\emph{Eq.(\ref{exercise}) }which however presumes the knowledge of the
Clifford calculus. We introduce the Dirac operator $\boldsymbol{\partial
=}\vartheta^{\mu}\mathbf{\nabla}_{\boldsymbol{\partial/\partial}x^{\mu}}$ and
recall that $dM=\boldsymbol{\partial}\wedge M$ for any $M\in\sec%
{\textstyle\bigwedge}
T^{\ast}M\hookrightarrow\sec\mathcal{C}\ell(M,\mathtt{g})$ where
$\mathcal{C}\ell(M,\mathtt{g})$ is the Clifford bundle of differential
forms\footnote{The Clifford product of Clifford fields is denoted by
justaposition.}. We have
\[
dH=d\underset{\boldsymbol{g}}{\star}G=\boldsymbol{\partial}\wedge
\underset{\boldsymbol{g}}{\star}G=-\underset{\boldsymbol{g}}{\star}J
\]
Now, we calculate $\boldsymbol{\partial}\wedge\underset{\boldsymbol{g}}{\star
}G=\boldsymbol{\partial}\wedge(\tilde{G}\boldsymbol{\tau}_{\boldsymbol{g}})$
in orthonormal basis \footnote{Recall that $\boldsymbol{\theta}^{\mathbf{a}%
}(\boldsymbol{e}_{\mathbf{b}})=\delta_{b}^{\mathbf{a}}$. Also $\boldsymbol{g}%
(\boldsymbol{e}_{\mathbf{a}},\boldsymbol{e}_{\mathbf{b}})=\eta_{\mathbf{ab}}$
and $\mathtt{g}(\boldsymbol{\theta}^{\mathbf{a}},\boldsymbol{\theta
}^{\mathbf{b}})=\eta^{\mathbf{ab}}=\boldsymbol{\theta}^{\mathbf{a}}%
\cdot\boldsymbol{\theta}^{\mathbf{b}}=\boldsymbol{\theta}^{\mathbf{a}%
}\lrcorner\boldsymbol{\theta}^{\mathbf{b}}$.}$\{\boldsymbol{e}_{\mathbf{a}}\}$
for $TU$ and\ dual basis $\{\boldsymbol{\theta}^{\mathbf{a}}\}$ for $T^{\ast
}U$ with $U\subset M$. We have
\begin{align*}
d\underset{\boldsymbol{g}}{\star}G  &  =\boldsymbol{\partial}\wedge(\tilde
{G}\boldsymbol{\tau}_{\boldsymbol{g}})\\
&  =-\boldsymbol{\theta}^{\mathbf{a}}\wedge(\mathbf{\nabla}_{\boldsymbol{e}%
_{\mathbf{a}}}(G\boldsymbol{\tau}_{\boldsymbol{g}}))\\
&  =-\frac{1}{2}\left(  \boldsymbol{\theta}^{\mathbf{a}}\mathbf{\nabla
}_{\boldsymbol{e}_{\mathbf{a}}}(G\boldsymbol{\tau}_{\boldsymbol{g}%
})+\mathbf{\nabla}_{\boldsymbol{e}_{\mathbf{a}}}(G\boldsymbol{\tau
}_{\boldsymbol{g}})\boldsymbol{\theta}^{\mathbf{a}}\right) \\
&  =-\frac{1}{2}\left(  \boldsymbol{\theta}^{\mathbf{a}}(\mathbf{\nabla
}_{\boldsymbol{e}_{\mathbf{a}}}G)\boldsymbol{\tau}_{\boldsymbol{g}%
}+\boldsymbol{\theta}^{\mathbf{a}}G(\mathbf{\nabla}_{\boldsymbol{e}%
_{\mathbf{a}}}\boldsymbol{\tau}_{\boldsymbol{g}})+(\mathbf{\nabla
}_{\boldsymbol{e}_{\mathbf{a}}}G)\boldsymbol{\tau}_{\boldsymbol{g}%
}\boldsymbol{\theta}^{\mathbf{a}}+G(\mathbf{\nabla}_{\boldsymbol{e}%
_{\mathbf{a}}}\boldsymbol{\tau}_{\boldsymbol{g}})\boldsymbol{\theta}%
^{a}\right) \\
&  =-\frac{1}{2}\left(  \boldsymbol{\theta}^{\mathbf{a}}(\mathbf{\nabla
}_{\boldsymbol{e}_{\mathbf{a}}}G)-(\mathbf{\nabla}_{\boldsymbol{e}%
_{\mathbf{a}}}G)\boldsymbol{\theta}^{\mathbf{a}})\boldsymbol{\tau
}_{\boldsymbol{g}}-\frac{1}{2}\boldsymbol{(}-\boldsymbol{\theta}^{\mathbf{a}%
}G(\mathbf{\nabla}_{\boldsymbol{e}_{\mathbf{a}}}\boldsymbol{\tau
}_{\boldsymbol{g}})-G(\mathbf{\nabla}_{\boldsymbol{e}_{\mathbf{a}}%
}\boldsymbol{\tau}_{\boldsymbol{g}})\boldsymbol{\theta}^{\mathbf{a}}\right)  .
\end{align*}
But%
\[
-\frac{1}{2}\left(  \boldsymbol{\theta}^{\mathbf{a}}(\mathbf{\nabla
}_{\boldsymbol{e}_{\mathbf{a}}}G)-(\mathbf{\nabla}_{\boldsymbol{e}%
_{\mathbf{a}}}G)\boldsymbol{\theta}^{\mathbf{a}})\boldsymbol{\tau
}_{\boldsymbol{g}}\right)  =-(\boldsymbol{\partial\lrcorner}G)\boldsymbol{\tau
}_{\boldsymbol{g}}%
\]
and
\[
\mathbf{\nabla}_{\boldsymbol{e}_{\mathbf{a}}}\boldsymbol{\tau}_{\boldsymbol{g}%
}=\mathbf{\nabla}_{\boldsymbol{e}_{\mathbf{a}}}(\boldsymbol{\theta
}^{\mathbf{0}}\wedge\boldsymbol{\theta}^{\mathbf{1}}\wedge\boldsymbol{\theta
}^{\mathbf{2}}\wedge\boldsymbol{\theta}^{\mathbf{3}})=-\Gamma_{\cdot
\mathbf{a0}}^{\mathbf{0\cdot\cdot}}\boldsymbol{\tau}_{\boldsymbol{g}}%
-\Gamma_{\cdot\mathbf{a1}}^{\mathbf{1\cdot\cdot}}\boldsymbol{\tau
}_{\boldsymbol{g}}-\Gamma_{\cdot\mathbf{a2}}^{\mathbf{2\cdot\cdot}%
}\boldsymbol{\tau}_{\boldsymbol{g}}-\Gamma_{\cdot\mathbf{a3}}^{\mathbf{3\cdot
\cdot}}\boldsymbol{\tau}_{\boldsymbol{g}}=0,
\]
because%
\[
\Gamma_{\cdot\mathbf{a0}}^{\mathbf{0\cdot\cdot}}=\Gamma_{\cdot\mathbf{a1}%
}^{\mathbf{1\cdot\cdot}}=\Gamma_{\cdot\mathbf{a1}}^{\mathbf{1\cdot\cdot}%
}=\Gamma_{\cdot\mathbf{a1}}^{\mathbf{1\cdot\cdot}}=0.
\]
Then,%
\[
-(\boldsymbol{\partial\lrcorner}G)\boldsymbol{\tau}_{\boldsymbol{g}%
}=-J\boldsymbol{\tau}_{\boldsymbol{g}}%
\]
or
\begin{equation}
\boldsymbol{\partial\lrcorner}G=J. \label{cinh}%
\end{equation}
Now,
\begin{align*}
\boldsymbol{\partial\lrcorner}G  &  =\frac{1}{2}\vartheta^{\mu}\lrcorner
(D_{\mu}G^{\alpha\beta}\vartheta_{\alpha}\wedge\vartheta_{\beta})=D_{\mu
}G^{\mu\beta}\vartheta_{\beta}\\
&  =(\partial_{\mu}G^{\mu\beta}+\Gamma_{\cdot\mu\kappa}^{\mu\cdot\cdot
}G^{\kappa\beta}+\Gamma_{\cdot\mu\kappa}^{\beta\cdot\cdot}G^{\mu\kappa
})\vartheta_{\beta}\\
&  =(\partial_{\mu}G^{\mu\beta}+\Gamma_{\cdot\mu\kappa}^{\mu\cdot\cdot
}G^{\kappa\beta})\vartheta_{\beta}=(\partial_{\mu}G^{\mu\beta}+\frac
{\partial_{\mu}\sqrt{\left\vert \det\boldsymbol{g}\right\vert }}%
{\sqrt{\left\vert \det\boldsymbol{g}\right\vert }}G^{\mu\beta})\vartheta
_{\beta}\\
&  =\frac{1}{\sqrt{\left\vert \det\boldsymbol{g}\right\vert }}\partial_{\mu
}(\sqrt{\left\vert \det\boldsymbol{g}\right\vert }G^{\mu\beta})\vartheta
_{\beta},
\end{align*}
and we get recalling that in vacuum $G=F$%
\[
\frac{1}{\sqrt{\left\vert \det\boldsymbol{g}\right\vert }}\partial_{\mu}%
(\sqrt{\left\vert \det\boldsymbol{g}\right\vert }F^{\mu\beta})=J^{\beta}%
\]
which is \emph{Eq.(\ref{exercise}) }again.
\end{remark}

\section{Maxwell Equations in Minkowski Spacetime}

\subsection{Vacuum Case}

We next suppose that electromagnetic phenomena take place in a non dispersive
material medium in Minkowski spacetime, i.e., the structure
$M=\langle\mathbb{R}^{4},\boldsymbol{\mathring{g}},\mathbf{\mathring{\nabla}%
},\boldsymbol{\tau}_{\boldsymbol{\mathring{g}}},\uparrow\rangle$. Since the
Riemann curvature of \ the Levi-Civita connection $\mathbf{\mathring{\nabla}}$
is null (i.e., $\boldsymbol{\mathring{R}}(\mathbf{\mathring{\nabla}})=0$)
there exists \emph{global }coordinates\footnote{These coordinates are said to
be in Einstein-Lorentz-Poincar\'{e} gauge. Note that $\langle\mathtt{x}^{\mu
}\rangle$ is a (ncsa%
$\vert$%
$\boldsymbol{e}_{0}$) where $\boldsymbol{e}_{0}=\boldsymbol{L}$ (the
laboratory frame) is an inertial reference system, this adjective meaning that
$\mathbf{\mathring{\nabla}}\boldsymbol{e}_{0}=0$.} $\langle\mathtt{x}^{\mu
}\rangle$ such that denoting by $\langle\boldsymbol{e}_{\mu}%
=\boldsymbol{\partial}/\boldsymbol{\partial}\mathtt{x}^{\mu}\rangle$ a basis
for $TM$ and $\langle\gamma^{\mu}=d\mathtt{x}^{\mu}\rangle$ the basis of
$T^{\ast}M$ dual to $\langle\boldsymbol{e}_{\mu}\rangle$ we have%
\begin{equation}
\boldsymbol{\mathring{g}}=\mathtt{\eta}_{\mu\nu}\gamma^{\mu}\otimes\gamma
^{\nu} \label{m19}%
\end{equation}
where the matrix with entries $\eta_{\mu\nu}$ being the diagonal matrix
\textrm{diag}$(1,-1,-1,-1)$.

In this case the components of constitutive (extensor) $\boldsymbol{\chi}$ of
the vacuum\ in the (nacs%
$\vert$%
$I$) $\{\mathtt{x}^{\mu}\}$\ to the inertial frame $\boldsymbol{L}%
=\boldsymbol{e}_{0}=\boldsymbol{\partial}/\boldsymbol{\partial}\mathtt{x}^{0}$ are%

\begin{equation}
\mathfrak{\chi}^{\rho\sigma\mu\nu}=(\mathtt{\eta}^{\rho\mu}\mathtt{\eta
}^{\sigma\nu}-\mathtt{\eta}^{\rho\nu}\mathtt{\eta}^{\sigma\mu}). \label{m20a}%
\end{equation}

Thus
\begin{equation}
\mathtt{G}^{\rho\sigma}=\frac{1}{2}(\mathtt{\eta}^{\rho\mu}\mathtt{\eta
}^{\sigma\nu}-\mathtt{\eta}^{\rho\nu}\mathtt{\eta}^{\sigma\mu})\mathring
{F}_{\mu\nu}=\mathtt{\eta}^{\rho\mu}\mathtt{\eta}^{\sigma\nu}\mathtt{F}%
_{\mu\nu}.
\end{equation}
Denoting by $(\mathtt{F}_{\mu\nu})$ and $(\mathtt{G}^{\rho\sigma})$,
respectively, the matrices with elements $\mathtt{F}_{\mu\nu}$ and
$\mathtt{G}^{\rho\sigma}$ we have
\begin{equation}
(\mathtt{F}_{\mu\nu})=\left(
\begin{array}
[c]{cccc}%
0 & \mathtt{E}_{1} & \mathtt{E}_{2} & \mathtt{E}_{3}\\
-\mathtt{E}_{1} & 0 & -\mathtt{B}_{3} & \mathtt{B}_{2}\\
-\mathtt{E}_{2} & \mathtt{B}_{3} & 0 & -\mathtt{B}_{1}\\
-\mathtt{E}_{3} & -\mathtt{B}_{2} & \mathtt{B}_{1} & 0
\end{array}
\right)  ,\qquad(\mathtt{G}^{\rho\sigma})=\left(
\begin{array}
[c]{cccc}%
0 & -\mathtt{B}_{1} & -\mathtt{B}_{2} & -\mathtt{B}_{3}\\
\mathtt{B}_{1} & 0 & \mathtt{E}_{3} & -\mathtt{E}_{2}\\
\mathtt{B}_{2} & -\mathtt{E}_{3} & 0 & \mathtt{E}_{1}\\
\mathtt{B}_{3} & \mathtt{E}_{2} & -\mathtt{E}_{1} & 0
\end{array}
\right)  .\nonumber
\end{equation}
Of course, Maxwell equations\ in coordinates in the Einstein-Lorentz
Poincar\'{e} gauge reads%

\begin{equation}
\frac{\partial\mathtt{F}_{\alpha\beta}}{\partial\mathtt{x}^{\mu}}%
+\frac{\partial\mathtt{F}_{\mu\alpha}}{\partial\mathtt{x}^{\beta}}%
+\frac{\partial\mathtt{F}_{\beta\mu}}{\partial\mathtt{x}^{\alpha}}=0,
\end{equation}
and%
\begin{equation}
\frac{\partial\mathtt{F}^{\mu\nu}}{\partial\mathtt{x}^{\mu}}=\mathtt{J}^{\nu}.
\end{equation}

\subsection{Non Dispersive Homogeneous and Isotropic Linear Medium Case}

We suppose in what follows that a linear non dispersive homogeneous and
isotropic medium (\textbf{NDHILM}) is at rest in a given reference frame
$\boldsymbol{V}$ in $U\subset M$ which has an \emph{arbitrary} motion relative
to the laboratory frame that is here modelled by the inertial frame
$\boldsymbol{L}=\boldsymbol{e}_{0}=\boldsymbol{\partial}/\boldsymbol{\partial
}\mathtt{x}^{0}$. Let $\langle x^{\prime\mu}\rangle$ be coordinate functions
covering $U$ that are (nacs%
$\vert$%
$\boldsymbol{V}$) and $\langle\boldsymbol{\partial/}\boldsymbol{\partial
}x^{\prime\mu}\rangle$ be a basis for $TU$ and $\langle\vartheta^{\prime\mu
}=dx^{\prime\mu}\rangle$ the corresponding dual basis for $T^{\ast}M$. Writing
in this case \
\begin{equation}
\boldsymbol{\mathring{g}}=\mathring{g}_{\mu\nu}^{\prime}\vartheta^{\prime\mu
}\otimes\vartheta^{\prime\nu}=\mathtt{\eta}_{\mu\nu}\gamma^{\mu}\wedge
\gamma\boldsymbol{\vartheta}^{\nu},
\end{equation}
we have
\begin{equation}
\boldsymbol{V}=V^{\prime\mu}\boldsymbol{\partial}/\boldsymbol{\partial
}x^{\prime\mu}=\frac{1}{\sqrt{\mathring{g}_{00}^{\prime}}}\boldsymbol{\partial
}/\boldsymbol{\partial}x^{\prime0}=\mathtt{V}^{\mu}\boldsymbol{\partial
}/\boldsymbol{\partial}\mathtt{x}^{\mu}=\mathtt{V}^{0}\frac
{\boldsymbol{\partial}}{\boldsymbol{\partial}\mathtt{x}^{0}}+\mathtt{V}%
^{i}\frac{\boldsymbol{\partial}}{\boldsymbol{\partial}\mathtt{x}^{i}}.
\label{m23}%
\end{equation}

We now write%
\begin{align}
F  &  =\frac{1}{2}F_{\rho\sigma}^{\prime}\vartheta^{\prime\rho}\wedge
\vartheta^{\prime\sigma}=\frac{1}{2}F^{\prime\rho\sigma}\vartheta_{\rho
}^{\prime}\wedge\vartheta_{\sigma}^{\prime}\nonumber\\
&  =\frac{1}{2}\mathtt{F}_{\rho\sigma}\gamma^{\rho}\wedge\gamma^{\sigma}%
=\frac{1}{2}\mathtt{F}^{\rho\sigma}\gamma_{\rho}\wedge\gamma_{\sigma
},\nonumber\\
G  &  =\frac{1}{2}G_{\rho\sigma}^{\prime}\vartheta^{\prime\rho}\wedge
\vartheta^{\prime\sigma}=\frac{1}{2}G^{\prime\rho\sigma}\vartheta_{\rho
}^{\prime}\wedge\vartheta_{\sigma}^{\prime}\nonumber\\
&  =\frac{1}{2}\mathtt{G}_{\rho\sigma}\gamma^{\rho}\wedge\gamma^{\sigma}%
=\frac{1}{2}\mathtt{G}^{\rho\sigma}\gamma_{\rho}\wedge\gamma_{\sigma}.
\label{m23a}%
\end{align}
We use the following notations for the elements of the matrices $(\mathtt{F}%
_{\rho\sigma}),(F_{\rho\sigma})$\ $(\mathtt{G}_{\rho\sigma})$ and
$(G_{\rho\sigma}^{\prime})$ as%
\begin{align}
(\mathtt{F}_{\rho\sigma})  &  =\left(
\begin{array}
[c]{cccc}%
0 & \mathtt{E}_{1} & \mathtt{E}_{2} & \mathtt{E}_{3}\\
-\mathtt{E}_{1} & 0 & -\mathtt{B}_{3} & \mathtt{B}_{2}\\
-\mathtt{E}_{2} & \mathtt{B}_{3} & 0 & -\mathtt{B}_{1}\\
-\mathtt{E}_{3} & -\mathtt{B}_{2} & \mathtt{B}_{1} & 0
\end{array}
\right)  ,\text{ \ \ }(F_{\rho\sigma}^{\prime})=\left(
\begin{array}
[c]{cccc}%
0 & E_{1}^{\prime} & E_{2}^{\prime} & E_{3}^{\prime}\\
-E_{1}^{\prime} & 0 & -B_{3}^{\prime} & B_{2}^{\prime}\\
-E_{2}^{\prime} & B_{3}^{\prime} & 0 & -B_{1}^{\prime}\\
-E_{3}^{\prime} & -B_{2}^{\prime} & B_{1}^{\prime} & 0
\end{array}
\right)  ,\label{m24}\\
(\mathtt{G}_{\rho\sigma})  &  =\left(
\begin{array}
[c]{cccc}%
0 & \mathtt{D}_{1} & \mathtt{D}_{2} & \mathtt{D}_{3}\\
-\mathtt{D}_{1} & 0 & -\mathtt{H}_{3} & \mathtt{H}_{2}\\
-\mathtt{D}_{2} & \mathtt{H}_{3} & 0 & -\mathtt{H}_{1}\\
-\mathtt{D}_{3} & -\mathtt{H}_{2} & \mathtt{H}_{1} & 0
\end{array}
\right)  ,\text{ \ \ }(G_{\rho\sigma}^{\prime})=\left(
\begin{array}
[c]{cccc}%
0 & D_{1}^{\prime} & D_{2}^{\prime} & D_{3}^{\prime}\\
-D_{1}^{\prime} & 0 & -H_{3}^{\prime} & H_{2}^{\prime}\\
-D_{2}^{\prime} & H_{3}^{\prime} & 0 & -H_{1}^{\prime}\\
-D_{3}^{\prime} & -H_{2}^{\prime} & H_{1}^{\prime} & 0
\end{array}
\right)  . \label{m24a}%
\end{align}
Then Maxwell equations read in the coordinates $\langle x^{\prime\mu}\rangle$
(which are a (nacs%
$\vert$%
$\boldsymbol{V}$)):%
\[
\frac{\partial F_{\alpha\beta}^{\prime}}{\partial x^{\prime\mu}}%
+\frac{\partial F_{\mu\alpha}^{\prime}}{\partial x^{\prime\beta}}%
+\frac{\partial F_{\beta\mu}^{\prime}}{\partial x^{\prime\alpha}}=0,
\]
{and}%
\[
\frac{1}{\sqrt{\left\vert \det\boldsymbol{\mathring{g}}\right\vert }}%
\frac{\partial}{\partial x^{\prime\alpha}}\left(  \frac{1}{\sqrt{\left\vert
\det\boldsymbol{\mathring{g}}\right\vert }}\mathring{g}^{^{\prime}\alpha\rho
}\mathring{g}^{\prime\beta\sigma}G_{\rho\sigma}^{\prime}\right)
=J^{\prime\beta}.
\]

\subsection{Minkowski Relations}

We want now to define using differential forms the concepts of electric and
magnetic fields and induction fields in a \textbf{NDHILM.}

Given an arbitrary reference frame $\boldsymbol{Z}$ in Minkowski spacetime let
$\boldsymbol{z=\mathring{g}(Z},)$ be the physically equivalent $1$-form
field\emph{\footnote{We will also call $\boldsymbol{z}$ a reference frame.}.}

\begin{definition}
The electric $\underset{\boldsymbol{z}}{\boldsymbol{E}}$ and the magnetic
$\underset{\boldsymbol{z}}{\boldsymbol{B}}$ $1$-form fields and the
$\underset{\boldsymbol{z}}{\boldsymbol{D}}$ and $\underset{\boldsymbol{z}%
}{\boldsymbol{H}}$ $1$-form fields according to the observers at rest in
$\boldsymbol{Z}$ are\emph{\footnote{The symbol $\star$ denotes the Hodge star
operator in Minkowski spacetime.}}:%

\begin{align}
\underset{\boldsymbol{z}}{\boldsymbol{E}}  &  :=\boldsymbol{z}\lrcorner
F,\text{ \ \ \ \ \ \ \ }\underset{\boldsymbol{z}}{\boldsymbol{B}%
}:=\boldsymbol{z}\lrcorner\star F,\label{eb}\\
\underset{\boldsymbol{z}}{\boldsymbol{D}}  &  :=\boldsymbol{z}\lrcorner
G,\text{ \ \ \ \ \ \ \ }\underset{\boldsymbol{z}}{\boldsymbol{H}%
}:=\boldsymbol{z}\lrcorner\star G. \label{hd0}%
\end{align}
We immediately have
\begin{align}
F  &  =\boldsymbol{z\wedge}\underset{\boldsymbol{z}}{\boldsymbol{E}}%
-\star(\boldsymbol{z}\wedge\underset{\boldsymbol{z}}{\boldsymbol{B}%
}),\label{eb1}\\
G  &  =\boldsymbol{z\wedge}\underset{\boldsymbol{z}}{\boldsymbol{D}}%
-\star(\boldsymbol{z}\wedge\underset{\boldsymbol{z}}{\boldsymbol{H}}).
\label{hd00}%
\end{align}

\end{definition}

Let $\boldsymbol{V}$ be an arbitrary reference frame in Minkowski spacetime
($\boldsymbol{v=\mathring{g}(V},)$) where the \textbf{NDHILM }is at rest.

\begin{definition}
\label{DHbis}The electric $\underset{\boldsymbol{v}}{\boldsymbol{D}}$ and the
magnetic $\underset{\boldsymbol{v}}{\boldsymbol{H}}$ $1$-form fields according
to the observers at rest in $\boldsymbol{Z}$ are related with the
$\underset{\boldsymbol{v}}{\boldsymbol{E}} \,\,{\mbox{and}}\,\,
\underset{\boldsymbol{v}}{\boldsymbol{B}}$ by:
\begin{equation}
\underset{\boldsymbol{v}}{\boldsymbol{D}}:=\boldsymbol{\varepsilon
}\underset{\boldsymbol{v}}{\boldsymbol{E}} \qquad\text{and} \qquad
\underset{\boldsymbol{v}}{\boldsymbol{H}}:=\frac{1}{\boldsymbol{\mu}%
}\underset{\boldsymbol{v}}{\boldsymbol{B}}. \label{dh}%
\end{equation}

\end{definition}

We immediately have
\begin{equation}
G=\frac{1}{\boldsymbol{\mu}}\left(  \boldsymbol{\varepsilon\mu}-1\right)
[\boldsymbol{v\wedge}(\boldsymbol{v}\lrcorner F)]+\frac{1}{\boldsymbol{\mu}}F.
\label{dh1}%
\end{equation}
which will be called \emph{Minkowski constitutive relation}.

\begin{exercise}
Prove \emph{Eq.(\ref{dh1}).}
\end{exercise}

\begin{solution}
From \emph{Eq.(\ref{hd00}) }and \emph{Definition \ref{DHbis}} we have
\begin{equation}
G=\varepsilon\boldsymbol{v}\wedge(\boldsymbol{v}\lrcorner F)-\frac{1}{\mu
}\star\lbrack\boldsymbol{v}\wedge(\boldsymbol{v}\lrcorner\star F)] \label{g1}%
\end{equation}
Now using the identities \emph{Eq.(\ref{tn2254})}\ and \emph{Eqs.(\ref{440new}%
)} in the \emph{Appendix} we have%
\begin{align*}
\star\lbrack\boldsymbol{v}\wedge(\boldsymbol{v}\lrcorner\star F)]  &
=\star\lbrack\boldsymbol{v}\wedge\star(\boldsymbol{v}\wedge F)]=\star
\star\lbrack\boldsymbol{v}\lrcorner(\boldsymbol{v}\wedge F)]=-\boldsymbol{v}%
\lrcorner(\boldsymbol{v}\wedge F)\\
&  =-(\boldsymbol{v}\lrcorner\boldsymbol{v)}F+\boldsymbol{v}\wedge
(\boldsymbol{v}\lrcorner F)
\end{align*}
and then%
\begin{align*}
G  &  =\varepsilon\boldsymbol{v}\wedge(\boldsymbol{v}\lrcorner F)+\frac{1}%
{\mu}[F-\boldsymbol{v}\wedge(\boldsymbol{v}\lrcorner F)]\\
&  =\frac{1}{\mu}\left(  \varepsilon\mu-1\right)  [\boldsymbol{v\wedge
}(\boldsymbol{v}\lrcorner F)]+\frac{1}{\mu}F.
\end{align*}

\end{solution}

\subsubsection{Coordinate Expression for Minkowski Constitutive Relations}

We want now to express Minkowski constitutive relations in arbitrary
$\{x^{\mu}\}$ coordinates covering $U\subset M$ and let $F_{\lambda\mu}$,
$G_{\mu\nu}$ and $v^{\mu}$ be the components of $F$, $G$ and the reference
frame $\boldsymbol{V}$ in an arbitrary natural bases $\{\boldsymbol{\partial
}_{\mu}=\boldsymbol{\partial/\partial}x^{\mu}\}$ and $\{\vartheta^{\mu
}=dx^{\mu}\}$. Let us calculate $\boldsymbol{v}\lrcorner G$. We have%
\[
\boldsymbol{v}\lrcorner G=\left(  \varepsilon-\frac{1}{\mu}\right)
\boldsymbol{v}\lrcorner\lbrack\boldsymbol{v\wedge}(\boldsymbol{v}\lrcorner
F)]+\frac{1}{\mu}\boldsymbol{v}\lrcorner F.
\]

From identity Eq.(\ref{tn2254}) in Appendix we have%
\begin{align}
\boldsymbol{v}\lrcorner\lbrack\boldsymbol{v\wedge}(\boldsymbol{v}\lrcorner F)]
&  =(\boldsymbol{v}\lrcorner\boldsymbol{v})(\boldsymbol{v}\lrcorner
F)]-\boldsymbol{v\wedge(v}\wedge(\boldsymbol{v}\lrcorner F))\nonumber\\
&  =\boldsymbol{v}\lrcorner F-(\boldsymbol{v\wedge v})\wedge(\boldsymbol{v}%
\lrcorner F))\nonumber\\
&  =\boldsymbol{v}\lrcorner F.\label{g20}%
\end{align}
Then%
\begin{equation}
\boldsymbol{v}\lrcorner G=\boldsymbol{\varepsilon v}\lrcorner F\label{g2}%
\end{equation}
and in components Eq.(\ref{g2}) is%
\[
v^{\mu}\vartheta_{\mu}\lrcorner\left(  \frac{1}{2}G_{\alpha\beta}%
\vartheta^{\alpha}\wedge\vartheta^{\beta}\right)  =\boldsymbol{\varepsilon
}v^{\mu}\vartheta_{\mu}\lrcorner\left(  \frac{1}{2}F_{\alpha\beta}%
\vartheta^{\alpha}\wedge\vartheta^{\beta}\right)
\]
and since $\vartheta_{\mu}\lrcorner\left(  \frac{1}{2}G_{\alpha\beta}%
\vartheta^{\alpha}\wedge\vartheta^{\beta}\right)  =G_{\mu\nu}\vartheta^{\nu}$,
$\vartheta_{\mu}\lrcorner\left(  \frac{1}{2}F_{\alpha\beta}\vartheta^{\alpha
}\wedge\vartheta^{\beta}\right)  =F_{\mu\nu}\vartheta^{\nu}$ we have%

\begin{equation}
v^{\mu}G_{\mu\nu}=\boldsymbol{\varepsilon}v^{\mu}F_{\mu\nu}. \label{g2a}%
\end{equation}

Let us now calculate $\boldsymbol{v}\lrcorner\star G$. First we note that%

\[
\star G=\left(  \boldsymbol{\varepsilon}-\frac{1}{\boldsymbol{\mu}}\right)
\star\lbrack\boldsymbol{v\wedge}(\boldsymbol{v}\lrcorner F)]+\frac
{1}{\boldsymbol{\mu}}\star F
\]
and using the same steps as the ones leading to Eq.(\ref{g2}) we first get%
\[
\boldsymbol{v}\lrcorner\star\lbrack\boldsymbol{v\wedge}(\boldsymbol{v}%
\lrcorner F)]=\star\lbrack\boldsymbol{v\wedge}(\boldsymbol{v}\wedge
(\boldsymbol{v\lrcorner}F))]=\star(\boldsymbol{v\wedge v)\wedge}%
(\boldsymbol{v\lrcorner}F)=0
\]
and then%
\begin{equation}
\boldsymbol{v}\lrcorner\star G=\frac{1}{\boldsymbol{\mu}}\boldsymbol{v}%
\lrcorner\star F.\label{gm2}%
\end{equation}
Now,
\begin{align*}
\star F &  =\frac{1}{2}\left(  \frac{1}{2}\sqrt{\left\vert \det
\boldsymbol{\mathring{g}}\right\vert }F^{\mu\nu}\varepsilon_{\mu\nu\rho\sigma
}\vartheta^{\rho}\wedge\vartheta^{\sigma}\right)  ,\\
\star G &  =\frac{1}{2}\left(  \frac{1}{2}\sqrt{\left\vert \det
\boldsymbol{\mathring{g}}\right\vert }G^{\mu\nu}\varepsilon_{\mu\nu\rho\sigma
}\vartheta^{\rho}\wedge\vartheta^{\sigma}\right)
\end{align*}
and
\begin{align*}
\boldsymbol{v}\lrcorner\star F &  =\frac{1}{2}\sqrt{\left\vert \det
\boldsymbol{\mathring{g}}\right\vert }v^{\kappa}F^{\mu\nu}\varepsilon_{\mu
\nu\kappa\sigma}\vartheta^{\sigma},\\
\boldsymbol{v}\lrcorner\star G &  =\frac{1}{2}\sqrt{\left\vert \det
\boldsymbol{\mathring{g}}\right\vert }v^{\kappa}G^{\mu\nu}\varepsilon_{\mu
\nu\kappa\sigma}\vartheta^{\sigma}%
\end{align*}
and finally Eq.(\ref{gm2}) reads in components%
\begin{equation}
\boldsymbol{\mu}v^{\kappa}G^{\mu\nu}\varepsilon_{\mu\nu\kappa\sigma}%
=v^{\kappa}F^{\mu\nu}\varepsilon_{\mu\nu\kappa\sigma}\label{GM2A}%
\end{equation}
or in equivalent forms%
\begin{align}
\boldsymbol{\mu}v_{\kappa}G_{\mu\nu}\varepsilon^{\mu\nu\kappa\sigma} &
=v_{\kappa}F_{\mu\nu}\varepsilon^{\mu\nu\kappa\sigma},\nonumber\\
\boldsymbol{\mu}\varepsilon^{\mu\nu\kappa\sigma}G_{\mu\nu}\mathring{g}%
_{\kappa\iota}v^{\iota} &  =\varepsilon^{\mu\nu\kappa\sigma}F_{\mu\nu
}\mathring{g}_{\kappa\iota}v^{\iota}.\label{GM2AA}%
\end{align}

\subsubsection{Polarization, Magnetization, Bound Current and Bound Charge
Fields}

We define moreover the \emph{polarization }$2-$\emph{form field }$\Pi$\emph{
}by%
\begin{equation}
\Pi:F-G. \label{POL}%
\end{equation}
\emph{Given an arbitrary frame }$\boldsymbol{Z}$ we decompose $\Pi$ as%
\begin{equation}
\Pi:=\boldsymbol{z\wedge}\underset{\boldsymbol{z}}{\boldsymbol{P}}%
-\star(\boldsymbol{z}\wedge\underset{\boldsymbol{z}}{\boldsymbol{M}}),
\label{POLA}%
\end{equation}
where $\underset{\boldsymbol{z}}{\boldsymbol{P}}$ and
$\underset{\boldsymbol{z}}{\boldsymbol{M}}$ are called respectively the
polarization $1$-form field and magnetization $1$-form field in
$\boldsymbol{Z}$. Moreover, from Eq.(\ref{dh}) and Eq.(\ref{dh1}) we get%
\[
\underset{\boldsymbol{z}}{\boldsymbol{P}}=\underset{\boldsymbol{z}%
}{\boldsymbol{E}}-\underset{\boldsymbol{z}}{\boldsymbol{D}}\text{
\ and\ \ }\underset{\boldsymbol{z}}{\boldsymbol{M}}=\underset{\boldsymbol{z}%
}{\boldsymbol{B}}-\underset{\boldsymbol{z}}{\boldsymbol{H}}.
\]

From the non homogeneous Maxwell equation $d\star G=-\mathbf{J}$ we have%
\begin{equation}
d\star F=-\mathbf{J}+d\star\Pi. \label{pol1}%
\end{equation}

The field
\begin{equation}
\mathcal{J}:=-d\star\Pi\label{pol2}%
\end{equation}
is called the \emph{bound current }$3$-form\emph{. }Given an arbitrary
frame\emph{ }$\boldsymbol{Z}$ we decompose $\mathcal{J}$ as%
\begin{equation}
\mathcal{J}:=\boldsymbol{z\wedge}\underset{\boldsymbol{z}}{\mathcal{J}%
}+\underset{\boldsymbol{z}}{\mathcal{\rho}} \label{pol3}%
\end{equation}
where $\underset{\boldsymbol{z}}{\mathcal{J}}$ is called the \emph{bound
current }$2$-form field and $\underset{\boldsymbol{z}}{\mathcal{\rho}}$ is
called the \emph{bound charge }$3$-form field\emph{.}

In the coordinates $\langle\mathtt{x}^{\mu}\rangle$ and $\langle x^{\prime\mu
}\rangle$ respectively adapted to the inertial laboratory frame
$\boldsymbol{L=e}_{\mathbf{0}}$ and to an arbitrary frame $\boldsymbol{V}$ we
write
\begin{align}
\Pi &  =\frac{1}{2}\mathtt{\Pi}_{\rho\sigma}\mathbf{\vartheta}^{\rho}%
\wedge\mathbf{\vartheta}^{\sigma}=\frac{1}{2}\mathtt{\Pi}^{\rho\sigma
}\mathbf{\vartheta}_{\rho}\wedge\mathbf{\vartheta}_{\sigma}\nonumber\\
&  =\frac{1}{2}\Pi_{\rho\sigma}^{\prime}\vartheta^{\prime\rho}\wedge
\vartheta^{\prime\sigma}=\frac{1}{2}\Pi^{\prime\rho\sigma}\vartheta_{\rho
}^{\prime}\wedge\vartheta_{\sigma}^{\prime}, \label{m27}%
\end{align}
where the entries of the matrices $(\mathring{\Pi}_{\rho\sigma})$ and
$(\Pi_{\rho\sigma})$ are, respectively,%
\begin{equation}
(\mathtt{\Pi}_{\rho\sigma}):=\left(
\begin{array}
[c]{cccc}%
0 & -\mathtt{P}_{1} & -\mathtt{P}_{2} & -\mathtt{P}_{3}\\
\mathtt{P}_{1} & 0 & -\mathtt{M}_{3} & \mathtt{M}_{2}\\
\mathtt{P}_{2} & \mathtt{M}_{3} & 0 & -\mathtt{M}_{1}\\
\mathtt{P}_{3} & -\mathtt{M}_{2} & \mathtt{M}_{1} & 0
\end{array}
\right)  ,\text{ \ \ }(\Pi_{\rho\sigma}):=\left(
\begin{array}
[c]{cccc}%
0 & -P_{1}^{\prime} & -P_{2}^{\prime} & -P_{3}^{\prime}\\
P_{1}^{\prime} & 0 & -M_{3}^{\prime} & M_{2}^{\prime}\\
P_{2}^{\prime} & M_{3}^{\prime} & 0 & -M_{1}^{\prime}\\
P_{3}^{\prime} & -M_{2}^{\prime} & M_{1}^{\prime} & 0
\end{array}
\right)  \label{m28}%
\end{equation}

\subsection{Constitutive Relations in a Uniformly Rotating Frame
$\boldsymbol{V}$.}

We introduce besides the coordinates $\{\mathtt{x}^{\mu}\}$ also
\emph{cylindrical coordinates} $\langle x^{\mu}\rangle$ in $\mathbb{R}%
^{4}-\{0\}$ which are also (nacs%
$\vert$%
$\boldsymbol{L}=\boldsymbol{e}_{0}$). As usual we write
\begin{align}
\mathtt{x}^{0}  &  =x^{0},\nonumber\\
\mathtt{x}^{1}  &  =x^{1}\cos x^{2},\nonumber\\
\mathtt{x}^{2}  &  =x^{1}\sin x^{2},\nonumber\\
\mathtt{x}^{3}  &  =x^{3}. \label{m29}%
\end{align}

We will simplify the notation when convenient by writing $\{\mathtt{x}%
^{0},\mathtt{x}^{1},\mathtt{x}^{2},\mathtt{x}^{3}\}:=\{t,\mathtt{x}%
,\mathtt{y},\mathtt{z}\}$ and $\{x^{0},x^{1},x^{2},x^{3}\}:=\{t,r,\phi,z\}$.

Next we introduce a particular \emph{rotating} reference frame $\boldsymbol{V}%
\in\sec TM$\ in Minkowski spacetime:%

\begin{equation}
\boldsymbol{V=}\frac{1}{\sqrt{1-\mathtt{v}^{2}}}\frac{\boldsymbol{\partial}%
}{\boldsymbol{\partial}t}+\frac{\omega}{\sqrt{1-\mathtt{v}^{2}}}\left(
-\mathtt{y}\frac{\boldsymbol{\partial}}{\boldsymbol{\partial}\mathtt{x}%
}+\mathtt{x}\frac{\boldsymbol{\partial}}{\boldsymbol{\partial}\mathtt{y}%
}\right)  \label{m30}%
\end{equation}
where
\begin{equation}
\mathtt{v}:=\sqrt{\mathtt{v}_{\mathtt{x}}^{2}+\mathtt{v}_{\mathtt{y}}^{2}%
}=\omega r, \label{m31}%
\end{equation}
with $\omega$ the (classical) angular velocity of $\boldsymbol{V}$ relative to
the inertial laboratory frame $\boldsymbol{L=e}_{0}$.

Since
\begin{equation}
\frac{\boldsymbol{\partial}}{\boldsymbol{\partial}\phi}=-y\frac
{\boldsymbol{\partial}}{\boldsymbol{\partial}x}+x\frac{\boldsymbol{\partial}%
}{\boldsymbol{\partial}y} \label{m32}%
\end{equation}
we can also write taking into account that $\gamma=1/\sqrt{1-\mathtt{v}^{2}}%
$,
\begin{align}
\boldsymbol{V}  &  =\frac{1}{\sqrt{1-\mathtt{v}^{2}}}\frac
{\boldsymbol{\partial}}{\boldsymbol{\partial}t}+\frac{\omega}{\sqrt
{1-\mathtt{v}^{2}}}\frac{\boldsymbol{\partial}}{\boldsymbol{\partial}\phi
}\nonumber\\
&  =\gamma\left(  \boldsymbol{\partial}_{t}+\mathtt{v}\frac{1}{r}%
\boldsymbol{\partial}_{\phi}\right)  \label{m33}%
\end{align}
and
\begin{equation}
\boldsymbol{v:=\mathring{g}}(\boldsymbol{V},\text{ })=\gamma\left(
dt+\mathtt{v}rd\phi\right)  . \label{m34a}%
\end{equation}

\begin{remark}
It is important to recall that although the coordinates $\{t,r,\phi,z\}$ cover
$\mathbb{R}^{4}-\{0\}$ the reference frame $\boldsymbol{V}$ if realized by a
physical system can only have material support for $r<1/\omega$.
\end{remark}

We introduce next the vector field
\begin{align}
\mathbf{v}  &  =-\omega\mathtt{y}\boldsymbol{\partial}_{\mathtt{x}}%
+\omega\mathtt{x}\boldsymbol{\partial}_{\mathtt{y}}\nonumber\\
&  =\mathtt{v}\frac{1}{r}\boldsymbol{\partial}_{\phi}=\mathtt{v}%
\boldsymbol{e}_{\phi} \label{m34}%
\end{align}

It is quite obvious that the vector field $\mathbf{v}$ represents the
classical $3$-\ velocity of a (material) point whose $3$-dimensional
trajectory in the spacelike section $\mathbb{R}^{3}$ determined by
$\boldsymbol{L=e}_{0}$ has parametric equations $(r\cos\omega t,r\sin\omega
t,z_{0})$ where $z_{0}$ is an arbitrary real constant.

Take notice that in engineering notation the vector field $\mathbf{v}$ is
usually denoted by $\vec{v}=\mathtt{v}\boldsymbol{\hat{e}}_{\phi}$. We will
use engineering notation when convenient in some of the formulas that follows
for pedagogical reasons.

\paragraph{A (nacs%
$\vert$%
$\boldsymbol{V}$)}

Before we proceed we introduce explicitly the transformation law between the
coordinates $\langle\mathtt{x}^{\mu}\rangle$ that are (nacs%
$\vert$%
$\boldsymbol{e}_{0}$) and $\langle x^{\prime\mu}\rangle$ that are (nacs%
$\vert$%
$\boldsymbol{V}$).

These are%
\begin{align}
\mathtt{x}^{0}  &  =x^{\prime0},\nonumber\\
\mathtt{x}^{1}  &  =x^{\prime1}\cos\omega\mathtt{x}^{0}-x^{\prime2}\sin
\omega\mathtt{x}^{0},\nonumber\\
\mathtt{x}^{2}  &  =x^{\prime1}\sin\omega\mathtt{x}^{0}+x^{\prime2}\cos
\omega\mathtt{x}^{0}.\nonumber\\
\mathtt{x}^{3}  &  =x^{\prime3}. \label{m35}%
\end{align}

To simplify the notation we will write $\{x^{\prime0},x^{\prime1},x^{\prime
2},x^{\prime3}):=(t^{\prime},x^{\prime},y^{\prime},z^{\prime})$.

We define also cylindrical coordinates $\langle\hat{x}^{\prime\mu}\rangle$
naturally adapted to $\boldsymbol{V}$ by%

\begin{align}
x^{\prime0}  &  =\hat{x}^{\prime0},\label{m36}\\
x^{\prime1}  &  =\hat{x}^{\prime1}\cos\hat{x}^{\prime2},\nonumber\\
x^{\prime2}  &  =\hat{x}^{\prime^{\prime}1}\sin\hat{x}^{\prime2},\nonumber\\
x^{\prime3}  &  =\hat{x}^{\prime3}.\nonumber
\end{align}

We moreover simply the notation writing \ $\{\hat{x}^{\prime0},\hat{x}%
^{\prime1},\hat{x}^{\prime2},\hat{x}^{\prime3}):=(t^{\prime},r^{\prime}%
,\phi^{\prime},z^{\prime}).$

Now, it is trivial to see that%
\begin{equation}
t^{\prime}=t,\text{ \ \ }r^{\prime}=r,\text{ \ \ }\phi^{\prime}=\phi-\omega
t,\text{ \ \ }z^{\prime}=\mathtt{z} . \label{m37}%
\end{equation}

This has as consequence the \ obvious relations%
\begin{equation}
dt^{\prime}=dt,\text{ \ \ }dr^{\prime}=dr,\text{ \ \ }d\phi^{\prime}%
=d\phi-\omega dt,\text{ \ \ }dz^{\prime}=d\mathtt{z} \label{m38}%
\end{equation}
and the not so obvious ones
\begin{equation}
\frac{\boldsymbol{\partial}}{\boldsymbol{\partial}t^{\prime}}=\frac
{\boldsymbol{\partial}}{\boldsymbol{\partial}t}+\omega\frac
{\boldsymbol{\partial}}{\boldsymbol{\partial}\phi},\text{ \ \ }\frac
{\boldsymbol{\partial}}{\boldsymbol{\partial}r^{\prime}}=\frac
{\boldsymbol{\partial}}{\boldsymbol{\partial}r},\text{ \ \ }\frac
{\boldsymbol{\partial}}{\boldsymbol{\partial}\phi^{\prime}}=\frac
{\boldsymbol{\partial}}{\boldsymbol{\partial}\phi},\text{ \ \ }\frac
{\boldsymbol{\partial}}{\boldsymbol{\partial}z^{\prime}}=\frac
{\boldsymbol{\partial}}{\boldsymbol{\partial}\mathtt{z}}. \label{m39}%
\end{equation}

Then we see that
\begin{equation}
\boldsymbol{V}=\frac{1}{\sqrt{1-\mathtt{v}^{2}}}\frac{\boldsymbol{\partial}%
}{\boldsymbol{\partial}t^{\prime}}=\frac{1}{\sqrt{\mathring{g}_{00}}}%
\frac{\boldsymbol{\partial}}{\boldsymbol{\partial}t^{\prime}}, \label{m40}%
\end{equation}
which shows that indeed $\langle\hat{x}^{\prime\mu}\rangle$ is a (nacs%
$\vert$%
$\boldsymbol{Z}$).

We also notice that since $r^{\prime}=r$ and $\frac{\boldsymbol{\partial}%
}{\boldsymbol{\partial}\phi^{\prime}}=\frac{\boldsymbol{\partial}%
}{\boldsymbol{\partial}\phi}$ we can write Eq.(\ref{m34}) as\footnote{Pay
attention with the notation used.
\par
{}}
\begin{align}
\mathbf{v}  &  =\omega\mathtt{v}\frac{1}{r}\frac{\boldsymbol{\partial}%
}{\boldsymbol{\partial}\phi}=\omega\mathtt{v}\frac{1}{r^{\prime}}%
\frac{\boldsymbol{\partial}}{\boldsymbol{\partial}\phi^{\prime}}=-\omega
y^{\prime}\frac{\boldsymbol{\partial}}{\boldsymbol{\partial}x^{\prime}}+\omega
x^{\prime}\frac{\boldsymbol{\partial}}{\boldsymbol{\partial}y^{\prime}%
}\nonumber\\
&  =v_{x^{\prime}}\frac{\boldsymbol{\partial}}{\boldsymbol{\partial}x^{\prime
}}+v_{y\prime}\frac{\boldsymbol{\partial}}{\boldsymbol{\partial}y^{\prime}%
}\label{m41}\\
&  =v_{1}^{\prime}\frac{\boldsymbol{\partial}}{\boldsymbol{\partial}%
x^{\prime1}}+v_{2}^{\prime}\frac{\boldsymbol{\partial}}{\boldsymbol{\partial
}x^{\prime2}}.\nonumber
\end{align}
We write the metric $\boldsymbol{\mathring{g}}$ using the coordinates $\langle
x^{\mu}\rangle$ and $\langle\hat{x}^{\mu}\rangle$ as%

\begin{align}
\boldsymbol{\mathring{g}}  &  =\mathring{g}_{\mu\nu}^{\prime}dx^{\prime\mu
}\otimes dx^{\prime\nu}\nonumber\\
&  =(1-\omega^{2}r^{\prime2})dt^{\prime}\otimes dt^{\prime}+2\omega y^{\prime
}dx^{\prime}\otimes dt^{\prime}-2\omega x^{\prime}dy\otimes dt^{\prime
}-dx^{\prime}\otimes dx^{\prime}-dy^{\prime}\otimes dy^{\prime}\nonumber\\
&  -dz^{\prime}\otimes dz^{\prime}, \label{metric1}%
\end{align}
or%
\begin{align}
\boldsymbol{\mathring{g}}  &  =\hat{g}_{\mu\nu}d\hat{x}^{\prime\mu}\otimes
d\hat{x}^{\prime\nu}\nonumber\\
&  =(1-\mathtt{v}^{2})dt\otimes dt-2\omega d\phi^{\prime}\otimes
dt-dr^{\prime}\otimes dr^{\prime}-r^{\prime2}d\phi^{\prime}\otimes
d\phi^{\prime}-dz^{\prime}\otimes dz^{\prime}. \label{metric2}%
\end{align}
We immediately read from Eq.(\ref{metric1}) that%
\begin{equation}
\mathring{g}_{00}^{\prime}=(1-\mathtt{v}^{2})=1/\gamma^{2},\text{
\ \ }\mathring{g}_{10}^{\prime}=-v_{1}^{\prime}=\omega y^{\prime},\text{
\ \ }\mathring{g}_{20}^{\prime}=-v_{2}^{\prime}=\omega x^{\prime},\text{
\ \ }\mathring{g}_{30}^{\prime}=0. \label{m43}%
\end{equation}

After this (long) digression we return to Eq.(\ref{g2a}) and Eq.(\ref{GM2AA})
that in coordinates $\langle x^{\prime\mu}\rangle$ which is (nacs%
$\vert$%
$\boldsymbol{V}$), can be immediately written in the\ engineering format (of
vector calculus) as:%

\begin{align}
\vec{D}^{\prime}  &  =\boldsymbol{\varepsilon}\vec{E}^{\prime},\label{m42}\\
\boldsymbol{\mu}[\frac{1}{\gamma^{2}}\vec{H}^{\prime}-\vec{v}\times\vec
{D}^{\prime}]  &  =\boldsymbol{[}\frac{1}{\gamma^{2}}\vec{B}^{\prime}-\vec
{v}\times\vec{E}^{\prime}].
\end{align}

Then, we finally get
\begin{gather}
\vec{D}^{\prime}=\boldsymbol{\varepsilon}\vec{E}^{\prime},\label{m44}\\
\vec{H}^{\prime}=\frac{1}{\boldsymbol{\mu}}[\vec{B}^{\prime}+\gamma
^{2}(\boldsymbol{\varepsilon\mu}-1)\vec{v}\times\vec{E}^{\prime}].
\label{m44a}%
\end{gather}

The \emph{polarization} and \emph{magnetization} vector fields in engineering
notation are, respectively, then:%
\begin{gather}
\vec{P}^{\prime}=(\boldsymbol{\varepsilon}-1)\vec{E}^{\prime},\label{M45}\\
\vec{M}^{\prime}=(1-\frac{1}{\boldsymbol{\mu}})\vec{B}^{\prime}+\frac
{1}{\boldsymbol{\mu}}\gamma^{2}(1-\boldsymbol{\varepsilon\mu})\vec{v}%
\times\vec{E}^{\prime}. \label{m45a}%
\end{gather}

To obtain the expression of those fields in the laboratory (in coordinates
$\langle\mathtt{x}^{\mu}\rangle$ naturally adapted to $\boldsymbol{l=e}%
_{0}=\boldsymbol{\partial}/\boldsymbol{\partial}\mathtt{x}^{0}%
=\boldsymbol{\partial}/\boldsymbol{\partial}t$) it is only necessary to recall
that
\begin{equation}
F_{\mu\nu}^{\prime}=\frac{\partial\mathtt{x}^{\alpha}}{\partial x^{\prime\mu}%
}\frac{\partial\mathtt{x}^{\beta}}{\partial x^{\prime\nu}}\mathtt{F}%
_{\alpha\beta},\text{ \ \ }H_{\mu\nu}^{\prime}=\frac{\partial\mathtt{x}%
^{\alpha}}{\partial x^{\prime\mu}}\frac{\partial\mathtt{x}^{\beta}}{\partial
x^{\prime\nu}}\mathtt{H}_{\alpha\beta},\text{ \ \ }\Pi_{\mu\nu}^{\prime}%
=\frac{\partial\mathtt{x}^{\alpha}}{\partial x^{\prime\mu}}\frac
{\partial\mathtt{x}^{\beta}}{\partial x^{\prime\nu}}\Pi_{\alpha\beta}.
\label{M46}%
\end{equation}

Now, from Eq.(\ref{m35}) we have%
\begin{equation}
\left(  \frac{\partial\mathtt{x}^{\alpha}}{\partial x^{\prime\mu}}\right)
=\left(
\begin{array}
[c]{cccc}%
1 & 0 & 0 & 0\\
-\omega\mathtt{y} & \cos\omega t^{\prime} & -\sin\omega t^{\prime} & 0\\
\omega\mathtt{x} & \sin\omega t^{\prime} & \cos\omega t^{\prime} & 0\\
0 & 0 & 0 & 1
\end{array}
\right)  . \label{jacobian}%
\end{equation}

Then, we have, e.g.,
\begin{align}
E_{x}^{\prime}  &  =\mathtt{E}_{\mathtt{x}}\cos\omega t+\mathtt{E}%
_{\mathtt{y}}\sin\omega t+\omega x^{\prime}\mathtt{B}_{\mathtt{z}},\nonumber\\
E_{x}^{\prime}  &  =-\mathtt{E}_{\mathtt{x}}\sin\omega t+\mathtt{E}%
_{\mathtt{y}}\cos\omega t+\omega y^{\prime}\mathtt{B}_{\mathtt{z}},\nonumber\\
E_{z}^{\prime}  &  =\mathtt{E}_{\mathtt{z}}-\omega\mathtt{yB}_{\mathtt{y}%
}-\omega\mathtt{xB}_{\mathtt{x}},\nonumber\\
B_{x}^{\prime}  &  =\mathtt{B}_{\mathtt{x}}\cos\omega t+\mathtt{B}%
_{\mathtt{y}}\sin\omega t,\nonumber\\
B_{y}^{\prime}  &  =-\mathtt{B}_{\mathtt{x}}\sin\omega t+\mathtt{B}%
_{\mathtt{y}}\cos\omega t,\nonumber\\
B_{z}^{\prime}  &  =\mathtt{B}_{\mathtt{z}}. \label{EE'}%
\end{align}

We recall also that writing at time $t$ in the laboratory the position vector
of a point $p=(\mathtt{x},\mathtt{y},\mathtt{z})$ in engineering notation as
$\mathtt{\vec{x}}=\mathtt{x}\boldsymbol{\hat{e}}_{\mathtt{x}}+\mathtt{y}%
\boldsymbol{\hat{e}}_{\mathtt{y}}+\mathtt{z}\boldsymbol{\hat{e}}_{\mathtt{z}}$
and the ($3$-dimensional) angular velocity \emph{field }of the frame
$\boldsymbol{V}$ as\ $\vec{\omega}=\Omega\boldsymbol{\hat{e}}_{\mathtt{z}}$ we
have%
\begin{equation}
\vec{v}=\vec{\Omega}\times\mathtt{\vec{x}.} \label{m49}%
\end{equation}
Finally we write the relation between the charge and current densities as
observed in the laboratory and the rotating frame. From $J=J^{\prime\mu
}\vartheta_{\mu}^{\prime}=J^{\mu}\gamma_{\mu}$, with $(J^{0},J^{1},J^{2}%
,J^{3})=(\rho,j_{x},j_{y},j_{z})$ and $(J^{^{\prime}0},J^{^{\prime}%
1},J^{^{\prime}2},J^{\prime3})=(\rho^{\prime},j_{x}^{\prime},j_{y}^{\prime
},j_{z}^{\prime})$ we have
\begin{equation}
J^{\prime\mu}=\frac{\partial x^{\prime\mu}}{\partial x^{\alpha}}J^{\alpha},
\label{m50}%
\end{equation}
which gives%
\begin{align}
\rho^{\prime}  &  =\mathtt{\rho},\label{m51}\\
j_{x}^{\prime}  &  =\omega y^{\prime}\mathtt{\rho}+\mathtt{j}_{\mathtt{x}}%
\cos\omega t+\mathtt{j}_{\mathtt{y}}\sin\omega t,\label{m51a}\\
j_{y}^{\prime}  &  =-\omega x^{\prime}\mathtt{\rho}-\mathtt{j}_{\mathtt{x}%
}\sin\omega t+\mathtt{j}_{\mathtt{y}}\cos\omega t,\label{m52b}\\
j_{z}^{\prime}  &  =\mathtt{j}_{\mathtt{z}}. \label{m52c}%
\end{align}

\section{Jump Conditions for the Fields $F$ and $G$ at the Boundary of a
Moving \textbf{NDHILM}}

The interface between a \emph{moving}\ \textbf{NDHILM} described by the
velocity field $\boldsymbol{V}$ and the vacuum defines a hypersurface
$\Sigma=0$ in Minkowski spacetime. The jump conditions can be deduced from
Maxwell equations and are well known. A \ very simple deduction of the
discontinuity of the fields $\vec{E},\vec{B},\vec{D}$ and $\vec{H}$ is given
in \cite{namias}. Here we write the jump conditions in differential form style
for the case where the free current $\star J=0$. Denoting as usual by $[F]$,
$[G]$ and $[\star G]$ the respective discontinuities of $F$, $G$ and $\star G$
at the boundary of medium and vacuum we have:%
\begin{gather}
\left.  \lbrack F]\wedge d\Xi\right\vert _{\Xi=0}=0,\label{J1}\\
\left.  \lbrack G]\llcorner d\Xi\right\vert _{\Xi=0}=0. \label{J2}%
\end{gather}

Observe that $\left.  [G]\llcorner d\Xi\right\vert _{\Xi=0}=0$ implies, of
course%
\[
\star\lbrack\left.  \lbrack G]\llcorner d\Xi\right\vert _{\Xi=0}%
]=-\star\lbrack\left.  \lbrack G]\lrcorner d\Xi\right\vert _{\Xi=0}]=\left.
[\star G]\wedge d\Xi\right\vert _{\Xi=0}=0.
\]
Also, $\left.  [F]\wedge d\Xi\right\vert _{\Xi=0}=$ $\left.  d\Xi\wedge\lbrack
F]\right\vert _{\Xi=0}=\left.  \star(d\Xi\lrcorner\star F)\right\vert _{\Xi
=0}=0$ and we can write the jump conditions also as:%

\begin{gather}
\left.  \lbrack\star F]\llcorner d\Xi\right\vert _{\Xi=0}=0,\label{JJ1}\\
\left.  \lbrack\star G]\wedge d\Xi\right\vert _{\Xi=0}=0. \label{JJ2}%
\end{gather}

Now, in the (nacs%
$\vert$%
$\boldsymbol{L=e}_{0}$) $\langle\mathtt{x}^{\mu}\rangle$ we have
\begin{equation}
n:=d\Xi=\frac{\partial\Xi}{\partial\mathtt{x}^{\mu}}\boldsymbol{\vartheta
}^{\mu}=n_{\mu}\boldsymbol{\vartheta}^{\mu}=n^{\mu}\boldsymbol{\vartheta}%
_{\mu}. \label{de}%
\end{equation}
Define\footnote{The minus sign is necessary due to the signature of the
metric.} $\mathbf{n}:=-\frac{\partial\Xi}{\partial\mathtt{x}^{i}}%
\frac{\boldsymbol{\partial}}{\boldsymbol{\partial}\mathtt{x}^{i}}$ the spatial
vector field \ which is normal to the moving boundary $\Xi(\mathtt{t}%
,\mathtt{x,y,z})=0$. Now, each spacial point of the moving boundary at time
$\mathtt{t}$ has Newtonian velocity $\mathbf{v}$. Observe that during a time
interval $\triangle\mathtt{x}^{0}=\triangle\mathtt{t}$ a point at the boundary
with arbitrary coordinates $(\mathtt{x},\mathtt{y},\mathtt{z})$ will arrive at
the point $(\mathtt{x}+v_{\mathtt{x}}\triangle\mathtt{t},\mathtt{y}%
+v_{\mathtt{y}}\triangle\mathtt{t},\mathtt{z}+v_{\mathtt{z}}\triangle
\mathtt{t})$. The hypersurface $\Xi$ \ at $(\mathtt{t}+\mathtt{\triangle
\mathtt{t}},\mathtt{x}+v_{\mathtt{x}}\triangle\mathtt{t},\mathtt{y}%
+v_{\mathtt{y}}\triangle\mathtt{t},\mathtt{z}+v_{\mathtt{z}}\triangle
\mathtt{t})$ will satisfy%
\begin{align}
&  \Xi(\mathtt{t}+\mathtt{\triangle\mathtt{t}},\mathtt{x}+v_{\mathtt{x}%
}\triangle\mathtt{t},\mathtt{y}+v_{\mathtt{y}}\triangle\mathtt{t}%
,\mathtt{z}+v_{\mathtt{z}}\triangle\mathtt{t})\nonumber\\
&  =\Xi(\mathtt{t},\mathtt{x,y,z})+\left.  \frac{\partial\Xi}{\partial
\mathtt{x}^{\mu}}\right\vert _{(\mathtt{t,x,y,z)}}\triangle\mathtt{x}^{\mu
}=0\nonumber\\
&  =\triangle\mathtt{t}\left.  \frac{\partial\Xi}{\partial\mathtt{t}%
}\right\vert _{(\mathtt{t,x,y,z)}}+v_{\mathtt{x}}\triangle\mathtt{t}\left.
\frac{\partial\Xi}{\partial\mathtt{x}}\right\vert _{(\mathtt{t,x,y,z)}%
}+v_{\mathtt{y}}\triangle\mathtt{t}\left.  \frac{\partial\Xi}{\partial
\mathtt{y}}\right\vert _{(\mathtt{t,x,y,z)}}+\left.  v_{\mathtt{z}}%
\triangle\mathtt{t}\frac{\partial\Xi}{\partial\mathtt{z}}\right\vert
_{(\mathtt{t,x,y,z)}}=0. \label{xx}%
\end{align}
Thus, we get in engineering notation taking into account that $\vec{n}%
=-\nabla\Xi$ that Eq.(\ref{xx}) implies
\begin{equation}
\frac{\partial\Xi}{\partial\mathtt{t}}-\vec{n}\bullet\vec{v}=0. \label{cont}%
\end{equation}
Then, we can write the jump conditions in its usual engineering notation as%
\begin{align}
\vec{n}\bullet\lbrack\underset{\boldsymbol{l}}{\vec{B}}]  &  =0,\text{
\ \ \ }(\vec{n}\bullet\vec{v})[\underset{\boldsymbol{l}}{\vec{B}}]+\vec
{n}\times\lbrack\underset{\boldsymbol{l}}{\vec{E}}]=0\label{vec1}\\
\vec{n}\bullet\lbrack\underset{\boldsymbol{l}}{\vec{D}}]  &  =0\text{,
\ \ \ }(\vec{n}\bullet\vec{v})[\underset{\boldsymbol{l}}{\vec{D}}]+\vec
{n}\times\lbrack\underset{\boldsymbol{l}}{\vec{H}}]=0. \label{vec2}%
\end{align}

\begin{exercise}
Show that \emph{Eq.(\ref{J2})} implies the formulas in \emph{Eq.(\ref{vec2}).}
\end{exercise}

\begin{solution}
We calculate $\left.  [G]\llcorner d\Xi\right\vert _{\Xi=0}=\left.
[G]\llcorner n\right\vert _{\Xi=0}=\left.  n\lrcorner\lbrack G]\right\vert
_{\Xi=0}=0$. Now, from \emph{Eq.(\ref{hd00})} we have%
\begin{equation}
n\lrcorner G=n\lrcorner(\boldsymbol{l}\wedge\underset{\boldsymbol{l}%
}{\boldsymbol{D}})-n\lrcorner\lbrack\star(\boldsymbol{i}\wedge
\underset{\boldsymbol{l}}{\boldsymbol{H}})] \label{vec3}%
\end{equation}
and since $\boldsymbol{l\lrcorner D=}\gamma^{0}\lrcorner(D_{i}\gamma^{i})=0$
we get
\begin{equation}
n\lrcorner(\boldsymbol{l}\wedge\underset{\boldsymbol{l}}{\boldsymbol{D}%
})=(n\lrcorner\boldsymbol{l)}\underset{\boldsymbol{l}}{\boldsymbol{D}}%
-n\wedge(\boldsymbol{l\lrcorner}\underset{\boldsymbol{l}}{\boldsymbol{D}%
})=n_{0}\underset{\boldsymbol{l}}{\boldsymbol{D}} \label{vec4}%
\end{equation}
which in engineering notation reads%
\begin{equation}
(\vec{n}\bullet\vec{v})\vec{D}. \label{vec5}%
\end{equation}
Also,
\begin{align}
n\lrcorner\lbrack\star(\boldsymbol{l}\wedge\boldsymbol{H})]  &  =\star
(n\wedge\boldsymbol{l}\wedge\boldsymbol{H})\nonumber\\
&  =\star(n_{i}H_{j}\gamma^{i}\wedge\gamma^{0}\wedge\gamma^{j})\nonumber\\
&  =-n_{i}H_{j}\star(\gamma^{0}\wedge\gamma^{i}\wedge\gamma^{j})\nonumber\\
&  =-n_{i}H_{j}\mathtt{\eta}^{0k}\mathtt{\eta}^{im}\mathtt{\eta}%
^{jn}\varepsilon_{kmnl}\gamma^{j}\nonumber\\
&  =-n_{i}H_{j}\mathtt{\eta}^{im}\mathtt{\eta}^{jn}\varepsilon_{0mnl}%
\gamma^{j}\nonumber\\
&  =-n^{m}H^{n}\varepsilon_{0mnl}\gamma^{j} \label{vec6}%
\end{align}
which in engineering notation read:%
\begin{equation}
-\vec{n}\times\vec{H}. \label{vec7}%
\end{equation}
Using \emph{Eq.(\ref{vec5})} and \emph{Eq.(\ref{vec7})} permit us to
write\ the equation $\left.  n\lrcorner\lbrack G]\right\vert _{\Xi=0}=0$ in
vector calculus notation as%
\[
(\vec{n}\bullet\vec{v})[\underset{\boldsymbol{l}}{\vec{D}}]+\vec{n}%
\times\lbrack\underset{\boldsymbol{l}}{\vec{H}}]=0.
\]
To show that $\left.  n\lrcorner\lbrack G]\right\vert _{\Xi=0}=0$ implies also
$\vec{n}\bullet\lbrack\underset{\boldsymbol{l}}{\vec{D}}]=0$ it is enough to
observe that
\begin{align}
\boldsymbol{l}\lrcorner(n\lrcorner\lbrack G])  &  =(\boldsymbol{l}\wedge
n)\lrcorner\lbrack G]\nonumber\\
&  =n_{i}(\gamma^{0}\wedge\gamma^{i})\lrcorner\lbrack\frac{1}{2}%
\mathtt{G}^{\alpha\beta}\gamma_{\alpha}\wedge\gamma_{\beta}]\nonumber\\
&  =-n_{i}(\gamma^{0}\wedge\gamma^{i})\cdot\lbrack\frac{1}{2}\mathtt{G}%
_{\alpha}^{\alpha\beta}\gamma\wedge\gamma_{\beta}]\nonumber\\
&  =-\frac{1}{2}n_{i}[\mathtt{G}^{\alpha\beta}]\det\left(
\begin{array}
[c]{cc}%
\gamma^{0}\cdot\gamma_{\alpha} & \gamma^{0}\cdot\gamma_{\beta}\\
\gamma^{i}\cdot\gamma_{\alpha} & \gamma^{i}\cdot\gamma_{\beta}%
\end{array}
\right) \nonumber\\
&  =-\frac{1}{2}n_{i}[\mathtt{G}^{\alpha\beta}](\delta_{\alpha}^{0}%
\delta_{\beta}^{i}-\delta_{\beta}^{0}\delta_{\alpha}^{i})\nonumber\\
&  =n_{i}[\mathtt{G}^{0i}]=0. \label{vec8}%
\end{align}
But $\mathtt{G}^{0i}=-\mathtt{D}^{i}$ and thus we can write $\boldsymbol{l}%
\lrcorner(n\lrcorner\lbrack G])=0$ in vector calculus notation as $\vec
{n}\bullet\lbrack\underset{\boldsymbol{l}}{\vec{D}}]=0$.
\end{solution}

\section{Solution of Maxwell Equations for the Wilson \& Wilson Experiment}

We now show how to find the solution of Maxwell equations
\begin{equation}
dF=0,\text{ \ \ \ }d\star G=-\mathbf{J} \label{magain}%
\end{equation}
for the famous Wilson \& Wilson experiment of 1913. Here that experiment is
modelled as follows: a cylindrical magnetic insulator of internal and external
radii $r_{1}$ and $r_{2}$ respectively and which has uniform and isotropic
electric and magnetic permeabilities $\mathbf{\varepsilon}$ and $\mathbf{\mu}$
is supposed to rotate with constant angular velocity $\mathbf{\omega}$ in the
$\mathtt{z}$ direction of an inertial laboratory $\boldsymbol{L=e}_{0}$ where
there exists a uniform magnetic field $F_{\mathbf{o}}=\mathrm{B}_{\mathbf{o}%
}dr\wedge d\phi$ (or in engineering notation $\vec{B}_{\mathbf{o}}%
=\mathrm{B}_{\mathbf{o}}\boldsymbol{\hat{e}}_{\mathtt{z}}$). Figure 1
illustrate the situation just described.%

\begin{figure}[ptb]%
\centering
\includegraphics[
natheight=11.694000in,
natwidth=8.263300in,
height=6.736in,
width=5.4034in
]%
{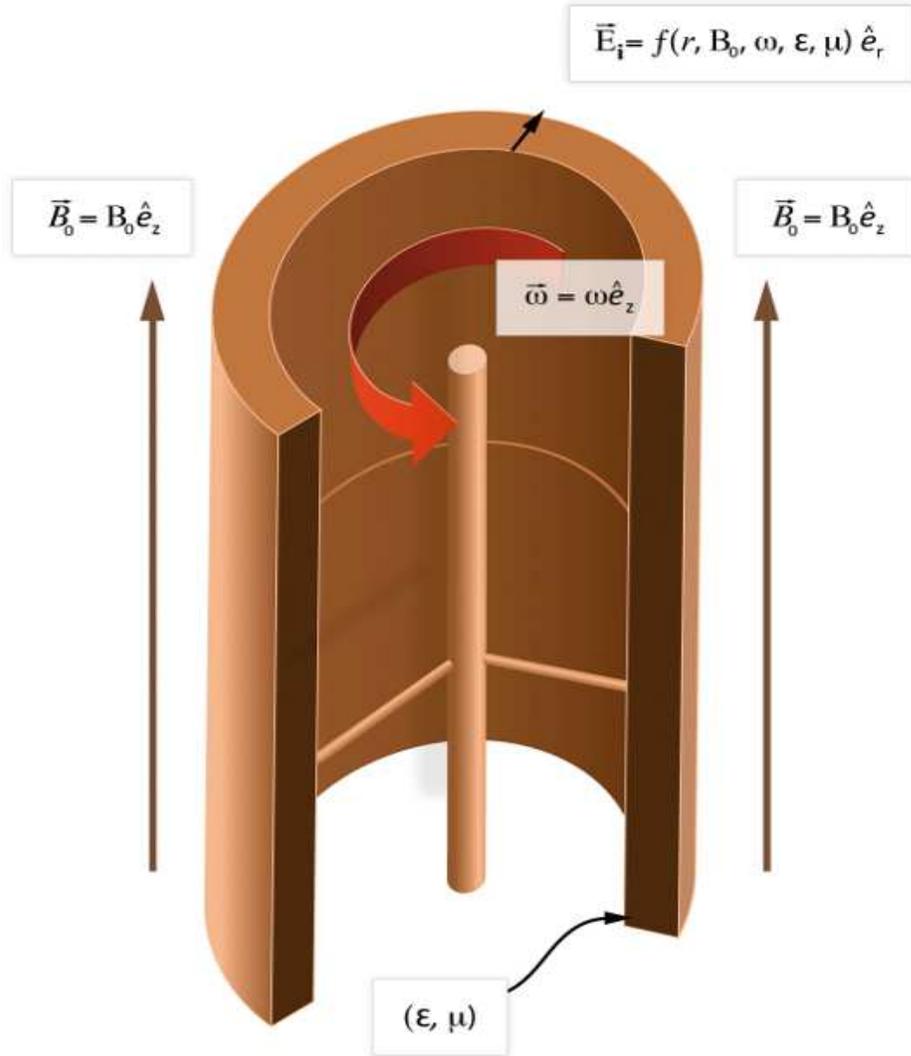}%
\caption{Details of the Wilson \& Wilson Experiment. A magnetic insulator
rotates with uniform angular velocity $\vec{\omega}=\omega\boldsymbol{\hat{e}%
}_{\mathtt{z}}$ in an external magnetic field\newline$\vec{B}_{\mathbf{o}%
}=\mathrm{B}_{\mathbf{o}}\boldsymbol{\hat{e}}_{\mathtt{z}}$. An electric field
$\vec{E}_{\mathbf{i}}=f(r,\mathrm{B}_{\mathbf{o}},\omega,\mathbf{\varepsilon
,\mu})\boldsymbol{\hat{e}}_{r}$ is observed.}%
\end{figure}

We start by introducing some useful notation. First \ we write the Minkowski
metric $\boldsymbol{\mathring{g}}$ as
\begin{equation}
\boldsymbol{\mathring{g}=\eta}_{\mu\nu}\boldsymbol{\theta}^{\mu}%
\otimes\boldsymbol{\theta}^{\nu} \label{gmin}%
\end{equation}
with%
\begin{equation}
\boldsymbol{\theta}^{0}=dt,\text{ \ \ }\boldsymbol{\theta}^{1}=dr,\text{
\ \ \ }\boldsymbol{\theta}^{2}=rd\phi,\text{ \ \ }\boldsymbol{\theta}^{3}=dz.
\label{tetas}%
\end{equation}

For our problem the moving boundary of our material has equations $\Xi
_{1}=r-r_{1}=0$ \ and $\Xi_{2}=r-r_{2}=0$, so the normal to this surface is
$d\Xi=\boldsymbol{\theta}^{1}$. For our problem we \ have $\mathbf{J}=0.$ To
proceed we write as an \emph{ansatz} the solution for the electromagnetic
field\ in the interior of the material%
\begin{equation}
F_{\mathbf{i}}=\mathcal{E}\boldsymbol{\theta}^{0}\wedge\boldsymbol{\theta}%
^{1}+\mathcal{B}\boldsymbol{\theta}^{1}\wedge\boldsymbol{\theta}^{2}
\label{fin}%
\end{equation}
where $\mathcal{E}$ and $\mathcal{B}$ are supposed to be functions only on the
coordinate $r$.

We recall from Eq.(\ref{m34a}) that the $1$-form physically equivalent to the
velocity field $\boldsymbol{V}$ is%
\[
\boldsymbol{v}=\gamma(\boldsymbol{\theta}^{0}+\mathtt{v}\boldsymbol{\theta
}^{2})\text{ }%
\]
with $\mathtt{v}=\omega r$ and $\gamma=(1-\mathtt{v}^{2})^{-1/2}$. We now must
solve the equation $d\star G=0$ with the boundary conditions given by
Eqs.(\ref{JJ1}) and (\ref{JJ2}). To calculate $G_{\mathbf{i}}$ we use
Eq.(\ref{dh1}). First we calculate $\boldsymbol{v\lrcorner}F_{\mathbf{i}}$. We
have
\begin{align*}
\boldsymbol{v\lrcorner}F_{\mathbf{i}}  &  =\gamma(\boldsymbol{\theta}%
^{0}+\mathtt{v}\boldsymbol{\theta}^{2})\lrcorner(\mathcal{E}\boldsymbol{\theta
}^{01}+\mathcal{B}\boldsymbol{\theta}^{12})\\
&  =\gamma(\mathcal{E}+\mathtt{v}\mathcal{B})\boldsymbol{\theta}^{1}.
\end{align*}
and%
\[
\boldsymbol{v}\wedge(\boldsymbol{v\lrcorner}F_{\mathbf{i}})=\gamma
^{2}(\mathcal{E}+\mathtt{v}\mathcal{B})\boldsymbol{\theta}^{01}-\gamma
^{2}\mathtt{v}(\mathcal{E}+\mathtt{v}\mathcal{B})\boldsymbol{\theta}^{12}.
\]
Then\footnote{It seems to have a misprint sign in the formula appearing in
\cite{catu}.}%
\begin{align}
G_{\mathbf{i}}  &  =\frac{\mathcal{E(\mathbf{\mu\varepsilon}-}\mathtt{v}%
^{2})+\mathcal{B\mathtt{v}(\mathbf{\mu\varepsilon}-}1)}{\mathcal{\mathbf{\mu}%
}(1-\mathtt{v}^{2})}\boldsymbol{\theta}^{01}+\frac{\mathcal{E\mathtt{v}%
(}1\mathcal{-\mathbf{\mu\varepsilon}})+\mathcal{B}(1-\mathcal{\mathbf{\mu
\varepsilon}}\mathtt{v}^{2})}{\mathcal{\mathbf{\mu}}(1-\mathtt{v}^{2}%
)}\boldsymbol{\theta}^{12}\nonumber\\
&  :=\mathcal{K}\boldsymbol{\theta}^{01}+\mathcal{L}\boldsymbol{\theta}^{12}.
\label{G}%
\end{align}

Since
\begin{align*}
\star(\boldsymbol{\theta}^{0}\wedge\boldsymbol{\theta}^{1})  &
=\widetilde{(\boldsymbol{\theta}^{0}\wedge\boldsymbol{\theta}^{1})}%
\lrcorner\boldsymbol{\tau}_{\boldsymbol{\mathring{g}}}=-\boldsymbol{\theta
}^{2}\wedge\boldsymbol{\theta}^{3},\\
\star(\boldsymbol{\theta}^{1}\wedge\boldsymbol{\theta}^{2})  &
=\widetilde{(\boldsymbol{\theta}^{1}\wedge\boldsymbol{\theta}^{2})}%
\lrcorner\boldsymbol{\tau}_{\boldsymbol{\mathring{g}}}=\boldsymbol{\theta}%
^{0}\wedge\boldsymbol{\theta}^{3}%
\end{align*}
we have
\begin{equation}
\star G_{\mathbf{i}}=-\mathcal{K}\boldsymbol{\theta}^{23}+\mathcal{L}%
\boldsymbol{\theta}^{03}. \label{dualG}%
\end{equation}
Then $d\star G=0$ gives%
\[
-\frac{d\mathcal{K}}{dr}\boldsymbol{\theta}^{123}-\frac{\mathcal{K}}%
{r}\boldsymbol{\theta}^{123}-\frac{d\mathcal{L}}{dr}\boldsymbol{\theta}%
^{013}=0
\]
and so our problem resumes in solving the following two trivial ordinary
differential equations
\begin{equation}
\frac{d\mathcal{K}}{dr}+\frac{\mathcal{K}}{r}=0\text{ \ and \ }\frac
{d\mathcal{L}}{dr}=0, \label{ode}%
\end{equation}
with solutions%
\begin{equation}
\mathcal{K}=\frac{c_{1}}{r}\text{ \ and \ }\mathcal{L}=c_{2} \label{sode}%
\end{equation}
respectively, where $c_{1}$ and $c_{2}$ are integration constants. So, we
have
\begin{align*}
G_{\mathbf{i}}  &  =\frac{c_{1}}{r}\boldsymbol{\theta}^{01}+c_{2}%
\boldsymbol{\theta}^{12},\\
\star G_{\mathbf{i}}  &  =-\frac{c_{1}}{r}\boldsymbol{\theta}^{23}%
+c_{2}\boldsymbol{\theta}^{03}.
\end{align*}
Now, if we recall that outside the magnetic insulator we have $\star
G_{\mathbf{o}}=\star F_{\mathbf{o}}=\mathrm{B}_{\mathbf{o}}\boldsymbol{\theta
}^{03}$ we get using the jump condition Eq.(\ref{JJ2}) that%
\begin{equation}
(\star G_{\mathbf{o}}-\star G_{\mathbf{i}})\wedge\boldsymbol{\theta}^{1}=0,
\end{equation}
i.e.,
\begin{equation}
-\mathrm{B}_{\mathbf{o}}\boldsymbol{\theta}^{013}+c_{2}\boldsymbol{\theta
}^{013}-\frac{c_{1}}{r}\boldsymbol{\theta}^{123}=0
\end{equation}
from where it follows that the functions $\mathcal{K}$ and $\mathcal{L}$ are
\begin{equation}
\mathcal{K}=0\text{, \ \ }\mathcal{L}=\mathrm{B}_{\mathbf{o}}. \label{KL}%
\end{equation}
Then from Eq.(\ref{G}) we have the following system of linear equations for
the functions $\mathcal{E}$ and $\mathcal{B}$:%

\begin{equation}
\left\{
\begin{array}
[c]{c}%
\mathcal{E(\mathbf{\mu\varepsilon}-}\mathtt{v}^{2})+\mathcal{B\mathtt{v}%
(\mathbf{\mu\varepsilon}-}1)=0,\\
\frac{\mathcal{E}\mathtt{v}\mathcal{(}1\mathcal{-\mathbf{\mu\varepsilon}%
})+\mathcal{B}(1-\mathcal{\mathbf{\mu\varepsilon}}\mathtt{v}^{2}%
)}{\mathcal{\mathbf{\mu}}(1-\mathtt{v}^{2})}=\mathrm{B}_{\mathbf{o}}%
\end{array}
\right.  \label{abeq}%
\end{equation}
whose solution is:%
\begin{equation}
\mathcal{E}=\mathrm{B}_{\mathbf{o}}\frac{\omega r(1-\mathcal{\mathbf{\mu
\varepsilon}})}{\mathcal{\mathbf{\varepsilon}}(1-\omega^{2}r^{2})},\text{
\ \ }\mathcal{B}=-\mathrm{B}_{\mathbf{o}}\frac{(\omega^{2}r^{2}%
-\mathcal{\mathbf{\mu\varepsilon}})}{\mathcal{\mathbf{\varepsilon}}%
(1-\omega^{2}r^{2})}. \label{solAB}%
\end{equation}

So, finally we have
\begin{equation}
F_{\mathbf{i}}=\mathrm{B}_{\mathbf{o}}\frac{\omega r(1-\mathcal{\mathbf{\mu
\varepsilon}})}{\mathcal{\mathbf{\varepsilon}}(1-\omega^{2}r^{2}%
)}\boldsymbol{\theta}^{01}-\mathrm{B}_{\mathbf{o}}\frac{(\omega^{2}%
r^{2}-\mathcal{\mathbf{\mu\varepsilon}})}{\mathcal{\mathbf{\varepsilon}%
}(1-\omega^{2}r^{2})}\boldsymbol{\theta}^{12}. \label{Fi}%
\end{equation}

and the electric and magnetic fields $\underset{\boldsymbol{l}}{\boldsymbol{E}%
_{\mathbf{i}}}$ \ and $\underset{\boldsymbol{l}}{\boldsymbol{H}_{\mathbf{i}}}$
as determined in the laboratory frame inside the material are%
\begin{align}
\underset{\boldsymbol{l}}{\boldsymbol{E}_{\mathbf{i}}}  &  =\boldsymbol{\theta
}^{0}\lrcorner F_{\mathbf{i}}=\mathrm{B}_{\mathbf{o}}\frac{\omega
r(1-\mathcal{\mathbf{\mu\varepsilon}})}{\mathcal{\mathbf{\varepsilon}%
}(1-\omega^{2}r^{2})}\boldsymbol{\theta}^{1},\label{TSO1}\\
\underset{\boldsymbol{l}}{\boldsymbol{H}_{\mathbf{i}}}  &  =\boldsymbol{\theta
}^{0}\lrcorner\star F_{\mathbf{i}}=-\mathrm{B}_{\mathbf{o}}\frac{(\omega
^{2}r^{2}-\mathcal{\mathbf{\mu\varepsilon}})}{\mathcal{\mathbf{\varepsilon}%
}(1-\omega^{2}r^{2})}\boldsymbol{\theta}^{3}. \label{TSO2}%
\end{align}
From Eq.(\ref{TSO1}) it follows immediately that the potential $V$ between the
internal and external parts of the material as shown in Figure 1 is, when
$\omega^{2}r^{2}<<1,$%
\begin{align}
V  &  =\mathrm{B}_{\mathbf{o}}\omega\frac{1}{\mathcal{\mathbf{\varepsilon}}%
}(1-\mathcal{\mathbf{\mu\varepsilon}})%
{\textstyle\int\nolimits_{r_{1}}^{r_{2}}}
rdr\nonumber\\
&  =\frac{1}{2}\mathrm{B}_{\mathbf{o}}\omega\frac{1}%
{\mathcal{\mathbf{\varepsilon}}}(1-\mathcal{\mathbf{\mu\varepsilon}}%
)(r_{2}^{2}-r_{1}^{2}), \label{V}%
\end{align}
which is the value found by Wilson \& Wilson \cite{ww}.

\begin{exercise}
Calculate $\underset{\boldsymbol{l}}{\boldsymbol{P}}$,
$\underset{\boldsymbol{l}}{\boldsymbol{M}}$, the bound current
$\underset{\boldsymbol{l}}{\mathcal{J}}$ and the bound charge
$\underset{\boldsymbol{l}}{\rho}$.
\end{exercise}

\section{Extracting Energy from the Magnetic Field}

We now propose a way to extract energy from a magnetic field using the results
just obtained above. In order to do that we first recall that when the
magnetic insulator in Figure 1 which has momentum of inertia $I$ is put in
rotation with constant angular velocity $\vec{\omega}=\omega\boldsymbol{\hat
{e}}_{\mathtt{z}}$ it acquires an angular momentum%
\begin{equation}
\vec{L}_{mec}=I\vec{\omega}. \label{e1}%
\end{equation}
For the preliminaries theoretical considerations in this section we suppose
that there are no energy losses due to friction of the rotating magnetic
insulator with its supporters nor losses due to Joule effect on electric
wires. As a consequence the \ total angular momentum of the system (i.e., the
rotating dielectric plus the electromagnetic field) must be conserved. We
suppose moreover that the system does not dissipate energy through radiation.
Under these conditions the total angular momentum of the system is
\begin{equation}
\vec{L}_{\mathbf{t}}=\vec{L}_{mec}+\vec{L}_{elec} \label{e2}%
\end{equation}
where by \cite{raruob} Abraham's formula\footnote{At first sight it may seems
strange that a magnetic and an electric field coming from different sources
may storage angular momentum. However that they do is an experimental fact as
showed for the first time by Graham and Lahoz only in 1979 \cite{grla,lagr}.}
\begin{equation}
\vec{L}_{elec}=%
{\textstyle\int}
d\mathtt{x}d\mathtt{y}d\mathtt{z}\overset{\rightarrow}{\mathtt{x}}\times
(\vec{E}_{\mathbf{i}}\times\vec{H}_{\mathbf{i}}). \label{e22}%
\end{equation}
We choose our coordinate system such that the origin lives in the middle of
the rotating axis of the dielectric. Under these conditions
\begin{align*}
\vec{E}_{\mathbf{i}}\times\vec{H}_{\mathbf{i}}  &  =-\mathrm{B}_{\mathbf{o}%
}^{2}\frac{\mathcal{\mathbf{\mu}}(1-\mathcal{\mathbf{\mu\varepsilon}})\omega
r(\omega^{2}r^{2}-\mathcal{\mathbf{\mu\varepsilon}})}%
{\mathcal{\mathbf{\varepsilon}}^{2}(1-\omega^{2}r^{2})^{2}}\boldsymbol{\hat
{e}}_{r}\times\hat{e}_{\mathtt{z}}\\
&  =f(r,\mathrm{B}_{\mathbf{o}},\omega,\mathcal{\mathbf{\mu}}%
,\mathcal{\mathbf{\varepsilon}})\boldsymbol{\hat{e}}_{\phi}.
\end{align*}
Then%
\begin{align*}
\mathtt{\vec{x}}\times(\vec{E}_{\mathbf{i}}\times\vec{H}_{\mathbf{i}})  &
=(\mathtt{z}\boldsymbol{\hat{e}}_{\mathtt{z}}+r\boldsymbol{\hat{e}}_{r})\times
f(r,\mathrm{B}_{\mathbf{o}},\omega,\mathcal{\mathbf{\mu}}%
,\mathcal{\mathbf{\varepsilon}})\boldsymbol{\hat{e}}_{\phi}\\
&  =\mathtt{z}f(r,\mathrm{B}_{\mathbf{o}},\omega,\mathcal{\mathbf{\mu}%
},\mathcal{\mathbf{\varepsilon}})\boldsymbol{\hat{e}}_{\mathtt{z}}%
\times\boldsymbol{\hat{e}}_{\phi}+rf(r,\mathrm{B}_{\mathbf{o}},\omega
,\mathcal{\mathbf{\mu}},\mathcal{\mathbf{\varepsilon}})\boldsymbol{\hat{e}%
}_{r}\times\boldsymbol{\hat{e}}_{\phi}\\
&  =-\mathtt{z}f(r,\mathrm{B}_{\mathbf{o}},\omega,\mathcal{\mathbf{\mu}%
},\mathcal{\mathbf{\varepsilon}})\boldsymbol{\hat{e}}_{r}+rf(r,\mathrm{B}%
_{\mathbf{o}},\omega,\mathcal{\mathbf{\mu}},\mathcal{\mathbf{\varepsilon}%
})\boldsymbol{\hat{e}}_{\mathtt{z}}\\
&  =-\mathtt{z}f(r,\mathrm{B}_{\mathbf{o}},\omega,\mathcal{\mathbf{\mu}%
},\mathcal{\mathbf{\varepsilon}})(\cos\phi\boldsymbol{\hat{e}}_{\mathtt{x}%
}+\sin\phi\boldsymbol{\hat{e}}_{\mathtt{y}})+rf(r,\mathrm{B}_{\mathbf{o}%
},\omega,\mathcal{\mathbf{\mu}},\mathcal{\mathbf{\varepsilon}}%
)\boldsymbol{\hat{e}}_{\mathtt{z}}.
\end{align*}
Thus the integral corresponding to the $\boldsymbol{e}_{r}$ component vanishes
and we have
\begin{align}
\vec{L}_{elec}  &  =\boldsymbol{e}_{\mathtt{z}}%
{\textstyle\int}
r^{2}f(r,\mathrm{B}_{\mathbf{o}},\omega,\mathcal{\mathbf{\mu}}%
,\mathcal{\mathbf{\varepsilon}})drd\phi d\mathtt{z}\nonumber\\
&  =2\pi Z\boldsymbol{e}_{\mathtt{z}}%
{\textstyle\int_{r^{1}}^{r_{2}}}
r^{2}f(r,\mathrm{B}_{\mathbf{o}},\omega,\mathcal{\mathbf{\mu}}%
,\mathcal{\mathbf{\varepsilon}})dr. \label{e3}%
\end{align}
where $Z$ is the height of the cylindrical dielectric. When $\omega^{2}%
r^{2}<<1$%
\begin{align}
\vec{L}_{elec}  &  =\boldsymbol{\hat{e}}_{\mathtt{z}}2\pi Z\frac
{\mathcal{\mathbf{\mu}}^{2}}{\mathcal{\mathbf{\varepsilon}}}\mathrm{B}%
_{\mathbf{o}}^{2}\omega(1-\mathcal{\mathbf{\mu\varepsilon}})%
{\textstyle\int_{r^{1}}^{r_{2}}}
r^{3}dr\nonumber\\
&  =-\frac{\pi Z\mathcal{\mathbf{\mu}}^{2}}{2\mathcal{\mathbf{\varepsilon}}%
}\mathrm{B}_{\mathbf{o}}^{2}\omega(\mathcal{\mathbf{\mu\varepsilon-}}1)\left(
r_{2}^{4}-r_{1}^{4}\right)  \boldsymbol{\hat{e}}_{\mathtt{z}}. \label{e4}%
\end{align}

It is a remarkable fact that $\vec{L}_{elec}$ points in the opposite direction
of $\vec{L}_{mec}$ if $\mathcal{\mathbf{\mu\varepsilon>}}1$ (as it is the case
in the Wilson \& Wilson experiment according to \cite{pelsw}. Since the total
angular momentum of our system must be conserved we see that $\left\vert
\vec{L}_{mec}\right\vert $ must increase if there is no energy losses due to
friction and Joule effect in the wires and we could use the potential $V$
given by Eq.(\ref{V}) to power an electric machine. Of course the energy
powering the electric machine can only be coming from the energy
\emph{stocked} in the external magnetic\ field. It will work while the
external magnetic field is presented. Of course extracting energy from the
magnetic field will make the value of $\mathrm{B}_{\mathbf{o}}$ to decrease
and with this decrease the potential $V$ will also decrease and at the end the
machine will stop working.

\begin{remark}
If $\mathcal{\mathbf{\mu\varepsilon<}}1$ the electromagnetic angular momentum
is in the same direction of the mechanical angular momentum, which must start
to decrease after the device is put in rotation. Then, the electric field will
decrease also making the electromagnetic momentum to decrease whereas the
mechanical angular momentum must then start to increase again. It is not clear
at this moment if the system will oscillate between a minimum and maximal
mechanical angular momentum or will stabilize.
\end{remark}

\begin{remark}
\textbf{Real World Machine}\emph{. }In the real world where friction and Joule
effect are always present in order to extract energy from the magnetic field
we need a machine like the one in Figure 2 where the potential difference $V$
given by \emph{Eq.(\ref{V})} is used first,\ if necessary, to power\ in a
compensator motor just enough to restore eventual losses due to friction and
Joule effect on the electric wires, and secondly $V$ is also used to power a
motor to generate useful work.
\end{remark}

\section{Conclusions}

In this paper we recalled how to correctly solve Maxwell equations in order to
find the electric field in the famous Wilson \& Wilson experiment using the
theory of differential forms with permits an intrinsic formulation of the
problem. We present also the correct jump conditions for the $F$ and $G$
fields in an invariant way\ when the boundary separating two media is in
motion. We give enough details for the paper to be useful for students and
researchers. Moreover we use the theoretical results to present a surprising
result which for the best of our knowledge is new: \emph{a machine that can
extract energy from an external magnetic field}. Under appropriate conditions
($\mathcal{\mathbf{\mu\varepsilon>}}1$) the machine once puts in motion will
start increasing the electromagnetic angular momentum stocked in the
electromagnetic field in the direction contrary to the original mechanical
angular momentum of the device which thus will start increasing
(theoretically) its angular speed until $\omega r_{2}=1$, thus generating a
big potential even in a small external magnetic field which can be use to
produce useful work.

Of course, a machine like this one if in orbit around a neutron star can
produce a lot of energy in a simple way than the hypothetical machine
described, e.g., in page 908 \cite{mtw} projected to extract energy from black
holes by throwing garbage on it.%

\begin{figure}[h]%
\centering
\includegraphics[
natheight=8.159500in,
natwidth=10.880200in,
height=6.5838in,
width=5.9257in
]%
{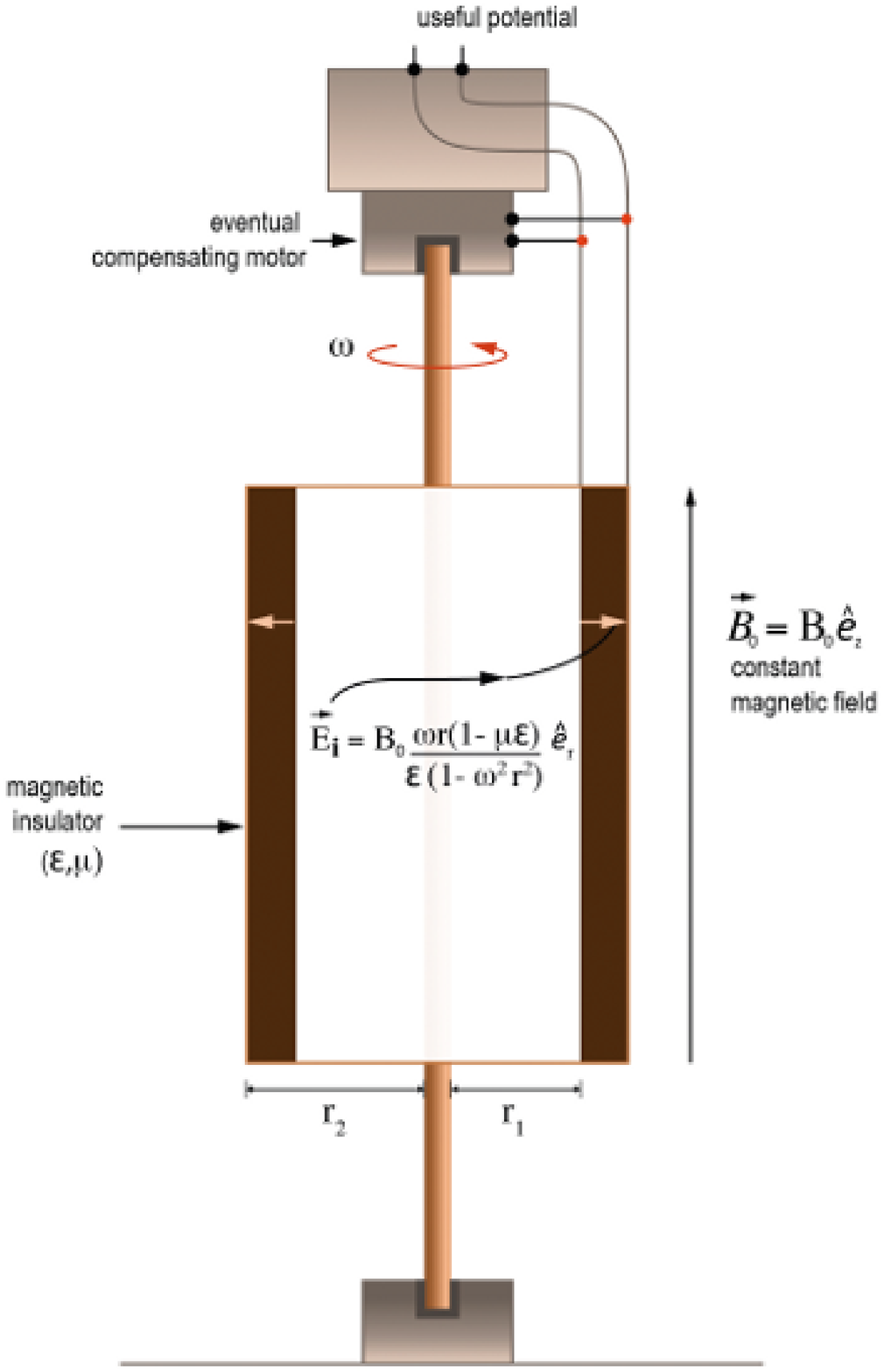}%
\caption{Machine to Extract Energy from an External Magnetic Field.}%
\end{figure}

\appendix{}

\section{Some Useful Formulas}

In this appendix we present some useful identities of the theory of
differential forms with has been used several times in the main text.

For $a\in\sec%
{\textstyle\bigwedge\nolimits^{1}}
T^{\ast}M$ and $A_{r}\in\sec%
{\textstyle\bigwedge\nolimits^{r}}
T^{\ast}M,B_{r}\in\sec%
{\textstyle\bigwedge\nolimits^{s}}
T^{\ast}M$ it is%

\begin{equation}
a\lrcorner(A_{r}\wedge B_{r})=(a\lrcorner A_{r})\wedge B_{r}+\hat{A}_{r}%
\wedge(a\lrcorner B_{r}). \label{tn2254}%
\end{equation}

\begin{equation}%
\begin{array}
[c]{l}%
A_{r}\wedge\underset{\boldsymbol{g}}{\star}B_{s}=B_{s}\wedge
\underset{\boldsymbol{g}}{\star}A_{r};\quad r=s\\
A_{r}\cdot\underset{\boldsymbol{g}}{\star}B_{s}=B_{s}\cdot
\underset{\boldsymbol{g}}{\star}A_{r};\quad r+s=n\\
A_{r}\wedge\underset{\boldsymbol{g}}{\star}B_{s}=(-1)^{r(s-1)}%
\underset{\boldsymbol{g}}{\star}(\tilde{A}_{r}\lrcorner B_{s});\quad r\leq s\\
A_{r}\lrcorner\underset{\boldsymbol{g}}{\star}B_{s}=(-1)^{rs}%
\underset{\boldsymbol{g}}{\star}(\tilde{A}_{r}\wedge B_{s});\quad r+s\leq n\\
\underset{\boldsymbol{g}}{\star}A_{r}=\tilde{A}_{r}\lrcorner\tau
_{\boldsymbol{g}}\\
\underset{\boldsymbol{g}}{\star}\tau_{\boldsymbol{g}}=\mathrm{sgn\det
}\boldsymbol{g};\quad\underset{\boldsymbol{g}}{\star}1=\tau_{\boldsymbol{g}}.
\end{array}
\label{440new}%
\end{equation}

\textbf{Acknowledgement }\emph{Authors are grateful to Flavia Tonelli for the
figures}.

\end{document}